\documentclass{aa}

\usepackage[tbtags,fleqn]{amsmath}
\usepackage{amsfonts}
\usepackage{pstricks, pst-plot}
\usepackage{graphics}
\usepackage{natbib}

\newcommand{\e}{\mathrm e}
\newcommand{\diff}{\mathrm d}

\newcommand{\R}{\mathbb R}
\newcommand{\mincir}{\raise
  -2.truept\hbox{\rlap{\hbox{$\sim$}}\raise5.truept \hbox{$<$}\ }}
\newcommand{\magcir}{\raise
  -2.truept\hbox{\rlap{\hbox{$\sim$}}\raise5.truept \hbox{$>$}\ }}
\newcommand{\nots}{\not\mskip 2.0mu s}


\DeclareMathOperator{\Cov}{Cov}

\begin{document}

\title{Interpolation and smoothing}
\author{Marco Lombardi}
\offprints{M. Lombardi}
\mail{lombardi@astro.uni-bonn.de}
\institute{%
  Institut f\"ur Astrophysik und Extraterrestrische Forschung,
  Universit\"at Bonn, Auf dem H\"ugel 71, D-53121 Bonn, Germany}
\date{Received ***date***; Accepted ***date***}
\abstract{%
  Smoothing is omnipresent in astronomy, because almost always
  measurements performed at discrete positions in the sky need to be
  interpolated into a smooth map for subsequent analysis.  Still, the
  statistical properties of different interpolation techniques are
  very poorly known.  In this paper, we consider the general problem
  of interpolating discrete data whose location measurements are
  distributed on the sky according to a known density distribution
  (with or without clustering).  We derive expressions for the
  expectation value and for the covariance of the smoothed map for
  many interpolation techniques, and obtain a general method that can
  be used to obtain these quantities for any linear smoothing.
  Moreover, we show that few basic properties of smoothing procedures
  have important consequences on the statistical properties of the
  smoothed map.  Our analysis allows one to obtain the statistical
  properties of an arbitrary interpolation procedure, and thus to
  optimally choose the technique that is most suitable for one's
  needs.  }

\maketitle

%

\defcitealias{2001A&A...373..359L}{Paper~I}
\defcitealias{MSCov}{Paper~II}

\section{Introduction}
\label{sec:introduction}

A common problem in Astronomy is the smoothing of irregularly sampled
data.  In general, this happens when one can make measurements on
discrete points of a quantity that has some astrophysical relevance.
In most cases, the ``quantity'' is a \textit{continuous field}, i.e.\ 
a smooth function on the sky.  In this case it is reasonable to try to
reconstruct the field by interpolating the discrete measurements
obtained.

Suppose, for example, that we are interested in measuring the column
density of a nearby molecular cloud.  A good estimate of the cloud
column density can be obtained, for example, using the infrared color
of background stars observed through the cloud \citep[see,
e.g.,][]{2001A&A...377.1023L}.  This way, we can obtain reliable
measurements of the ``column density field'' at the discrete points
corresponding to the star positions.  Finally, we interpolate the
various measurements and obtain a smooth map.  Similar situations are
often encountered in different fields (e.g., weak gravitational
lensing, peculiar velocity field of galaxies).

Interpolation and smoothing are ubiquitous in Astronomy, and indeed
many papers have been devoted to the study of the effects of
interpolation in particular analyses \citep[see,
e.g.][]{1992ApJ...398..169R}.  However, it is important to observe
that many different approaches to the study of interpolation are
possible, depending on the type of problem considered.  In this paper,
in particular, we will study the statistical properties of several
interpolation techniques.  Since the locations on the sky where
measurements are performed cannot generally be chosen in Astronomy
(cf.\ the example of reddening of stars above), we will carry out an
\textit{ensemble average\/} over the measurement locations.  More
precisely, we will assume that the measurements are randomly
distributed on the sky following a known scheme and we will carry out
a statistical analysis on this sample.  Note that there is a complete
freedom on the spatial distribution of measurements on the sky, that
can be correlated (for example, clustered) or can follow a non-uniform
density $\rho(x)$.  The ensemble average, which has already been
carried out under simplifying hypotheses of uncorrelated points and
uniform density in earlier papers \citetext{e.g.,
  \citealp{1998A&A...335....1L}; \citealp{2000MNRAS.313..524W};
  \citealp{2001A&A...373..359L}, hereafter Paper~I; \citealp{LPM};
  \citealp{MSCov}, hereafter Paper~II}, let us to derive general
results that generalize the particular configuration considered.

In previous works (see in particular Paper~I and II) we have focused
on a widely used smoothing method.  Here, in contrast, we will keep
the discussion much more general and consider some wide classes of
interpolating techniques.  This alternative approach has a number of
advantages: (i)~The results obtained are very general and can be
applied to several smoothing techniques (``proof reusability'');
(ii)~The analytic discussion is kept at a very simple level; (iii)~The
results obtained are more general, since allow for a non-uniform
density of locations and correlation on the locations; (iv)~The
properties of a smoothing technique can be predicted in advance
without the need of long calculations.  This last point, in our
opinion, is particularly important for practical applications.  In
fact, with the aid of the results obtained in this paper, several
smoothing techniques can be easily compared, which allows one to
choose the interpolation method more suitable.  We stress, however,
that some specific results cannot be derived using the techniques
described in this article and need a more specific analysis as done in
Paper~I and II.

The paper is organized in two main parts.  In the first part,
Sect.~\ref{sec:classifications}, we classify the smoothing methods and
we show the general results associated to each interpolation family.
In the second part, Sect.~\ref{sec:kern-some-smooth}, we illustrate
the results obtained earlier by considering several common smoothing
techniques.  Finally, in Sect.~\ref{sec:conclusions}, we briefly draw
the conclusions of this analysis.

\subsection{Notation}
\label{sec:notation}

Let us suppose to have a set of pairs $\bigl\{ (x_i, y_i) \bigr\}$,
where $x_i \in X$, called ``locations,'' belong to a real vector space
$X$ (typically, $X = \R^n$, with $n=1$, $2$, or $3$), and $y_i \in Y$,
called ``values,'' belong to a field $Y$ (in this paper, we will
assume for simplicity $Y = \R$).  The spatial interpolation problem
consists in finding a way to obtain approximated values for a generic
$x \in X$.  Formally, an interpolation procedure is a function $S$
that maps a point $x \in X$ and an \textit{unordered\/} set of couples
$(x_i, y_i) \in X \times Y$ into a value $y \in Y$:
\begin{equation}
  \label{eq:1}
  y = S\bigl( x; \bigl\{ (x_i, y_i) \bigr\} \bigr) \; .
\end{equation} 
Note that the number $N$ of couples $(x_i, y_i)$ is not fixed
\textit{a priori\/} (in particular $N$ could vanish).  In the
following we will use as synonymous interpolation procedure, smoothing
technique, interpolator.

As an example, let us consider a simple interpolator where the
smoothed value at $x$ is obtained as a weighted sum of the values $\{
y_i \}$.  More precisely, we define $S$ as
\begin{equation}
  \label{eq:2}
  S \bigl( x; \bigl\{ (x_i, y_i) \bigr\} \bigr) = \frac{\sum_{j=1}^N
  y_j / | x - x_j |^\alpha}{\sum_{j=1}^N 1 / | x - x_j |^\alpha} \; ,
\end{equation}
where $\alpha$ is a fixed real number.  This interpolation technique
will be often used in this paper to illustrate with an example some
general, abstract results.

In this paper we will study some statistical properties of various
interpolation procedures assuming that the locations $\{ x_i \}$ are
random variables distributed with spatial density $\rho(x)$ (with or
without clustering), and that the values $\{ y_i \}$ are associated to
the locations through the relation
\begin{equation}
  \label{eq:3}
  y_i = f(x_i) + \epsilon_i \; .
\end{equation}
Here $f \colon X \rightarrow Y$ is a known function and $\{ \epsilon_i
\}$, representing measurement errors, are random variables with
vanishing mean and covariance matrix (taking $\{ \epsilon_i \}$ as a
multidimensional random variable) proportional to the identity:
\begin{align}
  \label{eq:4}
  \langle \epsilon_i \rangle & {} = 0 \; , & 
  \langle \epsilon_i \epsilon_j \rangle & {} = \delta_{ij} \sigma^2 \; .
\end{align}
Note that the second relation of Eq.~\eqref{eq:4} states the so-called
``statistical orthogonality'' of the measurement errors; this property
is trivially satisfied if $\{ \epsilon_i \}$ are independent random
variables with fixed variance $\sigma^2$ (this is a good approximation
for many astronomical observations).  The subject of our study will be
the average value $\bigl\langle \tilde f(x) \bigr\rangle$ of $\tilde
f(x)$, where $\tilde f(x)$ is the interpolated value of $f$ at $x$:
\begin{equation}
  \label{eq:5}
  \tilde f(x) = S \bigl( x; \bigl\{ (x_i, y_i) \bigr\} \bigr) = 
  S \bigl( x; \bigl\{ (x_i, f(x_i) + \epsilon_i) \bigr\} \bigr) \; .
\end{equation}
We will also investigate the covariance (or two-point correlation) of
the map $\tilde f(x)$, defined as
\begin{align}
  \label{eq:6}
  \Cov(x, x'; \tilde f) & {} = \Bigl\langle \bigl[ \tilde f(x) -
  \bigl\langle \tilde f(x) \bigr\rangle \bigr] \bigl[ \tilde f(x') -
  \bigl\langle \tilde f(x') \bigr\rangle \bigr] \Bigr\rangle \notag\\
  & {} = \bigl\langle \tilde f(x) \tilde f(x') \bigr\rangle -
  \bigl\langle \tilde f(x) \bigr\rangle \bigl\langle \tilde f(x')
  \bigr\rangle\; .
\end{align}

A word of explanation is needed regarding averages.  Averages of
$\tilde f(x)$ and $\tilde f(x) \tilde f(x')$ are carried out both with
respect to $\{ \epsilon_i \}$ and to $\{ x_i \}$.  As we have
anticipated above, averages with respect to $\{ x_i \}$ are carried
out assuming that the locations are random variables distributed with
density $\rho(x)$ over the set $X$.  For example, if we assume that
the locations are uncorrelated and that the density is constant over
the field (this is usually referred to as a homogeneous Poisson
process or as compete spatial randomness), then the number $N$ of
object locations inside a field $A \subset X$ of finite area $\mu(A)$
follows a Poisson distribution with average $\rho \mu(A)$:
\begin{equation}
  \label{eq:7}
  p_N(N) = \e^{-\rho \mu(A)} \frac{ \bigl[\rho \mu(A) \bigr]^N}{N!} \;
  . 
\end{equation}
Under the same hypotheses, the $N$ locations $\{ x_i \}$ are, then,
uniformly distributed inside $A$.  In more general cases (in
particular, in case of correlation), it can be non-trivial to specify
exactly the probability distribution of points.  However, fortunately
for the following discussion we need only the density $\rho(x)$ and an
algorithm to randomly generate the locations (as discussed by
\citealp{1954ApJ...119...91S}; see also, e.g., the hierarchical model
described by \citealp{1978AJ.....83..845S}).  

The probability distribution for the other random variables considered
here, namely $\{ \epsilon_i \}$, need not to be fully specified.
Actually, for our purposes, we just need to specify that these
variables, representing measurement errors, have vanishing mean, fixed
variance, are orthogonal [i.e., satisfy Eq.~\eqref{eq:4}], and are
independent of the locations $\{ x_i \}$.  In the following
calculations, we will carry out first the average over the measurement
errors $\{ \epsilon_i \}$, and then the one over the locations $\{ x_i
\}$.

We finally note that in this study we allow for cases where the
smoothing procedure $S$ cannot be defined.  For example, if the
density $\rho$ is small, we might end up with a configuration without
locations $x_i$ close enough to $x$.  Some interpolation procedures
then might not be applied.  In the following, we adopt the convention
of discarding in the ensemble average the configurations $\{ x_i \}$
for which the interpolation procedure \eqref{eq:1} is not
\textit{locally\/} defined.  For example, when evaluating the average
$\bigl\langle \tilde f(x) \bigr\rangle$, we discard configurations
that make $S$ not defined at $x$; similarly, for expressions such as
$\bigl\langle \tilde f(x) \tilde f(x') \bigr\rangle$ we discard
configurations for which $S$ is not defined at $x$ or at $x'$.  Note
that we assume that the applicability of the smoothing technique
depends only on the locations $\{ x_i \}$ and \textit{not\/} on the
values $\{ y_i \}$.  We call $P_0(x)$ the probability of having a
configuration $\{ x_i \}$ for which the estimator is not defined at
$x$.  For example, some smoothing techniques (see
Sect.~\ref{sec:local-interpolators}) are not defined if there are no
locations inside a given subset $\pi$; in this case we find (in case
of vanishing correlation, i.e. for a Poisson distribution of locations)
\begin{equation}
  \label{eq:8}
  P_0(x) = \exp \biggl( -\int_\pi \rho(x) \, \diff x \biggr) \; .
\end{equation}
Similarly, we define $P_0(x, x')$ as the probability that the
smoothing is not defined either at $x$ or at $x'$ (or at both points).
If, again, the smoothing technique is defined at $x$ only if there is
at least a location inside a given subset $\pi$, and is defined at
$x'$ if there is a location inside the subset $\pi'$, then
\begin{align}
  \label{eq:9}
  & P_0(x,x') = \exp \biggl( - \int_{\pi \cap \pi'} \rho(x) \,
  \diff x \biggr) \notag\\
  & \qquad \times \biggl[ \exp \biggl( - \int_{\pi \setminus \pi'}
  \rho(x) \, \diff x \biggr) + \exp \biggl( - \int_{\pi' \setminus
    \pi} \rho(x) \, \diff x \biggr) \notag\\
  & \qquad \phantom{{} \times \biggl[} - \exp \biggl( - \int_{(\pi
    \setminus \pi') \cup (\pi' \setminus \pi)} \rho(x) \, \diff x
  \biggr) \biggr] \; .
\end{align}
Note that $P_0(x, x) = P_0(x)$ and that $P_0(x) P_0(x') \le P_0(x, x')
\le P_0(x) + P_0(x')$.  If an estimator is always defined, as for the
interpolator \eqref{eq:2}, we just set $P_0(x) = P_0(x, x') = 0$.

\section{Classification of interpolation procedures}
\label{sec:classifications}

Very little could be said at this stage about the average of $\tilde
f(x)$ or its covariance because the problem in the formulation of
Sect.~\ref{sec:notation} is by far too general.  However, as we will
see below, additional hypotheses allow us to obtain various results.
Some of these hypotheses are quite obvious, in the sense that we could
hardly imagine a sensible interpolation procedure that does not
satisfy them.  For example, if all values are the same, say $y_i = 1$,
we expect to have always $y = 1$ in Eq.~\eqref{eq:1} (this property is
discussed in Sect.~\ref{sec:normalization}).  Still this ``obvious''
hypothesis allows us to obtain some non-trivial results.

In this section we will consider some criteria used to classify
interpolation procedures, with the aim of introducing a standard
terminology and, at the same time, of deriving a number of useful
results.

\subsection{Linearity}
\label{sec:linearity}

Some of the most interesting interpolation procedures are linear
functions of the data values $\{ y_i \}$.  We stress that the
linearity is on the \textit{values\/} and not on the locations $\{ x_i
\}$.  Formally, the linearity is expressed by the relation
\begin{align}
  \label{eq:10}
  S \bigl( x; \bigl\{ (x_i, \alpha y_i + \beta y'_i) \bigr\}
  \bigr) = {} & \alpha S \bigl( x; \bigl\{ (x_i, y_i) \bigr\}
  \bigr) \notag\\
  & {} + \beta S \bigl( x; \bigl\{ (x_i, y'_i) \bigr\} \bigr) \; ,
\end{align}
where $\alpha$ and $\beta$ are arbitrary real numbers.  Given a
location configuration $\{ x_i \}$, we define the $j$-th
\textit{weight\/} of $S$ as
\begin{equation}
  \label{eq:11}
  w_j\bigl( x; \{ x_i \} \bigr) = S \bigl( x; \bigl\{ (x_i,
  \delta_{ij}) \bigr\} \bigr) \; .
\end{equation}
Using Eq.~\eqref{eq:11}, we can write any linear interpolation
procedure in the form
\begin{equation}
  \label{eq:12}
  S \bigl( x; \bigl\{ (x_i, y_i) \bigr\} \bigr) = \sum_{j=1}^N w_j
  \bigl( x; \{ x_i \} \bigr) y_j \; .
\end{equation}

For example, the smoothing technique \eqref{eq:2} is manifestly a
linear function of the values $\{ y_i \}$, and thus is a linear
interpolator.  Actually, it is already written in the form
\eqref{eq:12} with
\begin{equation}
  \label{eq:13}
  w_j \bigl( x; \{ x_i \} \bigr) = \frac{1 / |x -
  x_j|^\alpha}{\sum_{i=1}^N 1 / |x - x_i|^\alpha} \; .
\end{equation}

Since the discussion of linear smoothing techniques is quite lengthy,
we split it into two subsections, corresponding to the study of the
bias and the covariance of the interpolated map.

\subsubsection{Bias}
\label{sec:bias-linear}

Linearity is very convenient for a study of the bias of the smoothing,
since it allows us to write the average value of a smoothed function
$\tilde f$ as
\begin{equation}
  \label{eq:14}
  \bigl\langle \tilde f(x) \bigr\rangle = \int_X f(\bar x) K(x; \bar
  x) \rho(\bar x) \, \diff \bar x \; .
\end{equation}
In other words, the bias of the estimator is completely described by
the kernel function $K(x; \bar x)$.  Note also that the only
statistical property of measurement errors $\{ \epsilon_i \}$ that
have been used to derive Eq.~\eqref{eq:14} is $\langle \epsilon_i
\rangle = 0$; again, this is due to linearity.  Almost all
interpolators considered in this paper are linear.

In order to show Eq.~\eqref{eq:14}, we consider the average
$\bigl\langle \tilde f(x) \bigr\rangle$ and use the linearity relation
\eqref{eq:10}:
\begin{align}
  \label{eq:15}
  \bigl\langle \tilde f(x) \bigr\rangle = {} & \Bigl\langle S\bigl( x;
  \bigl\{ (x_i, f(x_i) + \epsilon_i) \bigr\} \bigr) \Bigr\rangle
  \notag\\
  {} = {} & \biggl\langle \sum_{j=1}^N \bigl[ f(x_j) + \epsilon_j
  \bigr] S\bigl( x; \bigl\{ (x_i, \delta_{ij}) \bigr\} \bigr)
  \biggr\rangle \; .
\end{align}
Here $\delta_{ij}$ is the Kronecker symbol.  We consider now the
average on errors, and note that the $\epsilon_j$ can be dropped from
the previous equation because $\langle \epsilon_j \rangle = 0$.  Let
us now decompose $f$ in the form
\begin{equation}
  \label{eq:16}
  f(x) = \int_X \delta(x - \bar x) f(\bar x) \, \diff \bar x \; ,
\end{equation}
where $\delta$ is Dirac's distribution.  Inserting this expression
into Eq.~\eqref{eq:15} we obtain
\begin{align}
  \label{eq:17}
  \bigl\langle \tilde f(x) \bigr\rangle = {} & \biggl\langle \int_X
  \diff \bar x \, f(\bar x) \sum_{j=1}^N \delta(x_j - \bar x) S\bigl(
  x; \bigl\{ (x_i, \delta_{ij}) \bigr\} \bigr) \biggr\rangle \notag\\
  {} = {} & \int_X f(\bar x) K(x; \bar x) \rho(\bar x) \, \diff \bar x
  \; ,
\end{align}
where $K(x; \bar x)$ is given by
\begin{align}
  \label{eq:18}
  K(x; \bar x) & {} = \frac{1}{\rho(\bar x)} \biggl\langle
  \sum_{j=1}^N \delta(x_j - \bar x) S\bigl( x; \bigl\{ (x_i,
  \delta_{ij}) \bigr\} \bigr) \biggr\rangle
  \notag\\
  & {} = \frac{1}{\rho(\bar x)} \Bigl\langle S\bigl( x; \bigl\{ (x_i,
  \delta(x_i - \bar x) \bigr\} \bigr) \Bigr\rangle \; .
\end{align}
Note that we have defined $K(x; \bar x)$ with a factor $1/\rho(\bar
x)$ for further convenience.  Equation~\eqref{eq:17} is precisely
Eq.~\eqref{eq:14}.

The practical evaluation of the kernel $K$ is very important for a
statistical study of the interpolation technique.  Indeed, this kernel
controls the relationship between the expected interpolated map and
the original, unknown field, and thus is useful, for example, to
perform a comparison between the observations and some theoretical
expectation.  In order to evaluate $K$, we first observe that almost
all linear interpolation procedures have a simple property: The
interpolated value at $x$ does not strongly depend on values that are
far away from it.  In this case, the kernel $K$ associated with a
linear interpolator can be obtained using a numerical technique that
we now describe (a proof will follow).  The procedure used in order to
determine $K(x; \bar x)$ is described in the following items:
\begin{enumerate}
\item Choose a large subset $A \subset X$ that contains both $x$ and
  $\bar x$.
\item Generate $N$, the number of objects inside $A$, according to the
  scheme chosen for the distribution of the locations (for example, if
  the locations are independent and uniformly distributed with density
  $\rho$, then $N$ follows the Poisson distribution given in
  Eq.~\eqref{eq:7}).
\item Generate the $N$ locations $\{ x_i \}$ inside $A$ following the
  scheme chosen (for example, uniformly if the density is constant and
  there is no correlation).
\item Assign a vanishing value $y_i = 0$ to each location $x_i$.
  Assign a value $1$ to an extra location at $\bar x$.
\item The configuration considered is composed by the object at $\bar
  x$ and the $N$ objects at $\{ x_i \}$; hence the configuration is
  $\Upsilon = \bigl\{ (\bar x, 1) \bigr\} \cup \bigl\{ (x_i, 0)
  \bigr\}$.
\item Evaluate $S(x; \Upsilon)$ if this function is defined at $x$;
  otherwise arbitrarily assign a vanishing value to the interpolated
  value at $x$.
\item Generate several configurations by repeating points 2--6, and
  evaluate the average of $S(x; \Upsilon)$.  This average, multiplied
  by $\bigl[ 1 - P_0(x) \bigr]^{-1}$, is an estimate of $K(x; \bar
  x)$.
\end{enumerate}
This procedure will be used in this paper not only to derive
numerically the kernels of several interpolating procedures, but also
to obtain some general properties of interpolators.

A proof of this practical method can be obtained as follows.  From
Eq.~\eqref{eq:18} we see that, in principle, $K$ could be evaluated by
performing an integration over the probability distribution function
for $\{ x_i \}$ using the whole set $X$.  In practice, this method
could never be applied in numerical studies because of the presence of
$\delta$ Dirac's distribution and because of the large number of
locations to be generated (actually, since $X$ is not bounded we
should generate an infinite number of objects).  However, a different
approach is feasible.  First, we use a Monte-Carlo integration, that
is, we generate a set of locations $\{ x_i \}$ according to the
expected probability distribution.  If we assume that the smoothing is
weakly dependent on values $y_i$ whose locations $x_i$ are far away
from $x$, we can safely generate points inside a subset $A \subset X$
that abundantly contains $x$ and $\bar x$.  Then, in order to solve
the problem with Dirac's delta, we approximate this distribution with
a top-hat function $\mathrm{H}(\bar x - x_i)$ centered on $\bar x$ and
with area $s$.  In other words, we take $\mathrm{H}(\bar x - x_i)$ to
be $1/s$ if $x_i$ falls inside a region of area $s$ around $\bar x$,
and we take $0$ otherwise; we will eventually let $s$ go to zero.
Since $s$ is taken to be small, in most cases all locations $\{ x_i
\}$ will not be close enough to $\bar x$ and thus all values $\{ y_i
\}$ will vanish.  In such situations, since the interpolator is
linear, we obtain a vanishing value at $x$.  In a few cases, however,
we expect to have configurations with a single point inside $s$, and
thus with value $1/s$; such configurations have probability $\rho(\bar
x) s$ as $s \rightarrow 0$.  When we now take the average of $S$,
cases where all values vanish do not contribute to the average because
the interpolated value vanishes as well; other cases, instead,
contribute with a term proportional to $1/s$ (the value corresponding
to the location around $\bar x$).  Equivalently, we can estimate the
relevant average forcing a point inside the top-hat close to $\bar x$
and multiplying the resulting average by the probability that this
happens, i.e.\ $\rho(\bar x) s$.  Since the interpolator is linear,
this is equivalent to forcing a point with value $1$ at $\bar x$,
taking the average, and multiplying the result by $\rho(\bar x)$; this
last factor, however, actually cancels with the term $1 / \rho(\bar
x)$ used in the definition of $K(x; \bar x)$ [cf. Eq.~\eqref{eq:18}].
This proves the correctness of the procedure to obtain $K$ described
in the points above.

In our discussion we have implicitly assumed that the estimator $S$ is
always defined, i.e.\ that $P_0(x) = 0$.  If this is not the case, we
must slightly modify the method described above.  In particular, we
need to distinguish between four probabilities: $P_{s,1}$, the
probability of having a point inside $s$ and the smoothing procedure
defined at $x$, $P_{\nots,1}$, the probability of having no point
inside $s$ and the smoothing procedure defined at $x$, and the two
symmetrical probabilities $P_{s,0}$ and $P_{\nots, 0}$ for
configurations where the interpolator is not defined at $x$.  Note
that $P_{s,0} + P_{\nots, 0} = P_0(x)$ as defined in
Sect.~\ref{sec:notation}.  We then find that the procedure described
above can be still applied if $P_0(x) \neq 0$, provided that two
changes are made:
\begin{itemize}
\item We need to discard, in the ensemble average of points,
  configurations for which $S$ cannot be evaluated at $x$.
\item We need to multiply the average of $S$ obtained from Monte-Carlo
  integration by the probability that, in the ensemble average, a
  point falls inside $s$.  In the general case $P_0(x) \neq 0$
  considered here, this probability is given by $P_{s,1} / \bigl[ s
  (P_{s,1} + P_{\nots,1}) \bigr]$.
\end{itemize}
Both problems can be solved at the same time using the following
procedure.  First, we note that the if we set $y = 0$ for cases where
$S$ cannot be evaluated at $x$, we are basically multiplying the
average by a factor $P_{s,1} / ( P_{s,0} + P_{s,1} )$.  On the other
hand, we know from the previous analysis that $P_{s,0} + P_{s,1} =
\rho(\bar x) s$.  Hence, we can recover the correct factor by setting
$y=0$ if $S$ is not defined and by multiplying the final result by
$\bigl[ 1 - P_0(x) \bigr]^{-1}$.  This completes the proof for $K(x;
\bar x)$.

\subsubsection{Covariance}
\label{sec:covariance-linear}

Linearity has also important consequences for the form of the
covariance.  In particular, the term $\bigl\langle \tilde f(x) \tilde
f(x') \bigr\rangle$ can be written as
\begin{align}
  \label{eq:19}
  & \bigl\langle \tilde f(x) \tilde f(x') \bigr\rangle = \int_X \diff
  \bar x \, \rho(\bar x) C_1(x, x'; \bar x) \bigl[ \sigma^2 + f(\bar
  x) f(\bar x) \bigr] \notag\\
  & \quad {} + \int_X \diff \bar x \, \rho(\bar x) \int_X \diff
  \bar x' \, \rho(\bar x') C_2(x, x'; \bar x, \bar x') f(\bar x)
  f(\bar x') \; .
\end{align}
Hence, the covariance is composed by a term proportional to
$\sigma^2$, the scatter of measurements $\{ y_i \}$ around the true
values $\bigl\{ f(x_i) \bigr\}$, and additional terms that can be
identified as Poisson noise.  More specifically, the two noise terms
are [cf.\ Eq.~\eqref{eq:6}]
\begin{align}
  \label{eq:20}
  & T_\sigma(x, x') = \sigma^2 \int_X \diff \bar x \, \rho(\bar x)
  C_1(x, x'; \bar x) \; , \\
  \label{eq:21}
  & T_\mathrm{P}(x, x') = \int_X \diff \bar x \, \rho(\bar x)
  C_1(x, x'; \bar x) \bigl[ f(\bar x) \bigr]^2 \notag\\
  & \quad {} + \int_X \diff \bar x \, \rho(\bar x) \int_X \diff \bar
  x' \, \rho(\bar x') C_2(x, x'; \bar x, \bar x') f(\bar x) f(\bar x')
  \notag\\
  & {} \quad {} - \int_X \diff \bar x \, \rho(\bar x) \int_X \diff
  \bar x' \, \rho(\bar x') K(x; \bar x) K(x'; \bar x') f(\bar
  x) f(\bar x') \; .
\end{align}
Again, the fact that measurement errors enter only through $\sigma^2$
is a consequence of linearity.  

In order to show Eq.~\eqref{eq:19}, we use again the linearity of $S$
to write [cf.\ Eq.~\eqref{eq:15}]
\begin{align}
  \label{eq:22}
  \tilde f(x) & {} = S\bigl( x; \bigl\{ (x_i, f(x_i) + \epsilon_i)
  \bigr\} \bigr) \notag\\
  & {} = \sum_{j=1}^N \bigl[ f(x_j) + \epsilon_j \bigr] S\bigl( x;
  \bigl\{ (x_i, \delta_{ij}) \bigr\} \bigr) \; .
\end{align}
Using this expression in the product $\bigl\langle \tilde f(x) \tilde
f(x') \bigr\rangle$, we obtain
\begin{align}
  \label{eq:23}
  \bigl\langle \tilde f(x) \tilde f(x') \bigr\rangle = {} &
  \biggl\langle \sum_{j, j'} \bigl[ f(x_j) + \epsilon_j \bigr] \bigl[
  f(x_{j'}) + \epsilon_{j'} \bigr] \notag\\
  & \phantom{\biggl\langle} \times S \bigl( x; \bigl\{ (x_i,
  \delta_{ij}) \bigr\} \bigr) S \bigl( x'; \bigl\{ (x_{i'},
  \delta_{i'j'}) \bigr\} \bigr)
  \biggr\rangle \notag\\
  {} = {} & \biggl\langle \sum_j \bigl[ f(x_j) f(x_j) + \sigma^2
  \bigr] S \bigl( x; \bigl\{ (x_i, \delta_{ij}) \bigr\}
  \bigr) \notag\\
  & \phantom{\biggl\langle} \times S \bigl( x'; \bigl\{ (x_{i'},
  \delta_{i'j}) \bigr\} \bigr) \biggr\rangle \notag\\
  & {} + \biggl\langle \sum_{j \neq j'} f(x_j) f(x_{j'}) S \bigl( x;
  \bigl\{ (x_i, \delta_{ij}) \bigr\}
  \bigr) \notag\\
  & \phantom{{} + \biggl\langle} \times S \bigl( x'; \bigl\{ (x_{i'},
  \delta_{i'j'}) \bigr\} \bigr) \biggr\rangle
\end{align}
Note that we have explicitly separated the cases $j = j'$ and $j \neq
j'$ in order to take advantage of Eq.~\eqref{eq:4}.  We now use again
the decomposition \eqref{eq:16} for $f(x)$, $\bigl[ f(x) \bigr]^2$,
and $\sigma^2$ (the last taken to be a constant function of $x$), thus
obtaining
\begin{align}
  \label{eq:24}
  & \bigl\langle \tilde f(x) \tilde f(x') \bigr\rangle = \int \diff
  \bar x \, \rho(\bar x) \bigl[ f(\bar x) f(\bar x) + \sigma^2 \bigr]
  C_1(x, x'; \bar x) \notag\\
  & \quad {} + \int \diff \bar x \, \rho(\bar x) \int \diff \bar x' \,
  \rho(\bar x') f(\bar x) f(\bar x') C_2(x, x'; \bar x, \bar x') \; ,
\end{align}
with
\begin{align}
  \label{eq:25}
  & C_1(x, x'; \bar x) = \frac{1}{\rho(\bar x)} \biggl\langle
  \sum_j \delta(\bar x - x_j) S\bigl( x; \bigl\{ (x_i, \delta_{ij})
  \bigr\} \bigr) \notag\\ 
  & \quad {} \times S \bigl( x'; \bigl\{ (x_{i'}, \delta_{i'j})
  \bigr\}\bigr) \biggr\rangle \; , \\
  \label{eq:26}
  & C_2(x, x'; \bar x, \bar x') = \frac{1}{\rho(\bar x) \rho(\bar x')}
  \biggl\langle \sum_{j \neq j'} \delta(\bar x - x_j) \delta(\bar x' -
  x_{j'}) \notag\\
  & \quad {} \times S\bigl( x; \bigl\{ (x_i, \delta_{ij}) \bigr\}
  \bigr) S \bigl( x'; \bigl\{ (x_{i'}, \delta_{i'j'}) \bigr\}\bigr)
  \biggr\rangle \; .
\end{align}
Again, the factors $\rho(\bar x)$ and $\rho(\bar x')$ have been
introduced here to simplify some of the following equations.  This
proves Eq.~\eqref{eq:19}.  Note that, although Eq.~\eqref{eq:26} is
apparently composed of two independent factors, $S\bigl( x; \bigl\{
(x_i, \delta_{ij}) \bigr\} \bigr)$ and $S \bigl( x'; \bigl\{ (x_{i'},
\delta_{i'j'}) \bigr\}\bigr)$, in reality both terms are functions of
\textit{all\/} locations $\{ x_i \}$; hence, since the random
variables that enters the expression for $C_2$ are precisely the
locations $\{ x_i \}$, the two terms are correlated and
Eq.~\eqref{eq:26} cannot further simplified.  The same, clearly,
applies to Eq.~\eqref{eq:25}.

Similarly to the average, the two kernels $C_1$ and $C_2$ can be
numerically evaluated using a simple procedure that we now describe:
\begin{enumerate}
\item Choose a large subset $A \subset X$ that contains $x$,
  $\bar x$, $x'$, and $\bar x'$.
\item Generate $N$, the number of objects inside $A$, according to the
  scheme chosen for the distribution of the locations (for example, if
  the locations are independent and uniformly distributed with density
  $\rho$, then $N$ follows the Poisson distribution given in
  Eq.~\eqref{eq:7}).
\item Generate the $N$ locations $\{ x_i \}$ inside $A$ following the
  scheme chosen (for example, uniformly if the density is constant and
  there is no correlation).
\item Assign a vanishing value $y_i = 0$ to each location $x_i$.  For
  $C_1$, assign a value $1$ to an extra location at $\bar x$,
  similarly to what we do for $K$.  For $C_2$, assign a value $1$ to
  two extra locations at $\bar x$ and $\bar x'$.
\item Hence, the two configurations considered are $\Upsilon_1 =
  \Upsilon = \bigl\{ (\bar x, 1) \bigr\} \cup \bigl\{ (x_i, 0)
  \bigr\}$ and $\Upsilon_2 = \bigl\{ (\bar x, 1), (\bar x', 1) \bigr\}
  \cup \bigl\{ (x_i, 0) \bigr\}$
\item For $C_1$ evaluate $S(x; \Upsilon_1) S(x'; \Upsilon_1)$, if the
  function $S$ is defined at both points $x$ and $x'$; otherwise
  arbitrarily assign a vanishing value to the product above.
  Analogously, for $C_2$ evaluate $S(x; \Upsilon_2) S(x';
  \Upsilon_2)$, if this product is defined, or use zero otherwise.
\item Repeat points 2--6, and evaluate the average $\bigl\langle S(x;
  \Upsilon_1) S(x'; \Upsilon_1) \bigr\rangle$.  This average,
  multiplied by $\bigl[ 1 - P_0(x, x') \bigr]^{-1}$, is an estimate of
  $C_1(x, x'; \bar x)$.  An estimate of $C_2(x, x'; \bar x, \bar x')$
  is instead given by the average $\bigl\langle S(x; \Upsilon_2) S(x';
  \Upsilon_2) \bigr\rangle$ multiplied by $\bigl[ 1 - P_0(x, x')
  \bigr]^{-1}$.
\end{enumerate}
Again, the numerical techniques described in the items above will be
also used to derive some general properties for the covariance of
interpolators in the following sections.

This practical method to evaluate $C_1$ and $C_2$ can be derived using
a technique similar to the one described for $K$.  In particular,
Eq.~\eqref{eq:25} suggests that we can determine the kernel $C_1$ as
follows.  We approximate the delta distribution in $\bar x$ with a
top-hat function $\mathrm{H}(\bar x - x_j)$ of area $s$.  Then, we
randomly generate the locations inside a large subset $A \subset X$
that contains $x$, $x'$, and $\bar x$.  As for the kernel $K$, most
configurations will have only points outside the top-hat, thus
resulting in a vanishing sum.  A fraction $s \rho(\bar x)$ of
configurations, however, will have a single point falling inside the
top-hat function, i.e.\ with coordinate very close to $\bar x$.  Since
the configurations without points inside $s$ do not contribute to the
average, we can just ignore them and consider only configurations with
a point at $\bar x$; we then multiply the average by the probability
of having such configurations, i.e.\ $\rho(\bar x) s$ in the limit $s
\rightarrow 0$.  Since the function $\mathrm{H}$ has a value $1/s$ at
$\bar x$, this is equivalent to assigning a value $1$ to the point at
$\bar x$, and multiply the final result by $\rho(\bar x)$; this term,
however, disappears because of the presence of a factor $1 / \rho(\bar
x)$ in the definition of $C_1$.

Again, in this procedure, we have ignored cases where the smoothing
technique cannot be applied.  In order to evaluate the expression in
the right hand side of Eq.~\eqref{eq:25}, we need the values of $S$ at
$x$ \textit{and\/} $x'$.  Hence, we should discard cases where $S$
cannot be evaluated at $x$ or $x'$.  Alternatively, we can proceed as
done for $K$, and just assign, for the configuration $\Upsilon$,
vanishing value to the product $S(x; \Upsilon) S(x'; \Upsilon)$ when
one of the two factors cannot be evaluated.  Finally, we multiply the
total result by a factor $\bigl[ 1 - P_0(x, x') \bigr]^{-1}$, where
this extra term is used to correctly normalize the average of
Eq.~\eqref{eq:25}.

The evaluation for $C_2$ is similar.  In this case, however, we have
two different delta distributions at $\bar x$ and $\bar x'$.  Hence,
we approximate them with two top-hat functions $\mathrm{H}(\bar x -
x_j)$ and $\mathrm{H}(\bar x' - x'_j)$, both of area $s$.  Since we
are going to take the limit $s \rightarrow 0$, the two top-hat
functions do not intersect, and thus the probability of having a point
inside both top-hats is the product of the individual probabilities,
i.e.\ $s^2 \rho(\bar x) \rho(\bar x')$.  Then, in the limit $s
\rightarrow 0$, we can force a point at $\bar x$ and one at $\bar x'$,
both with value $1$; other points are randomly distributed with
density $\rho$.  Then, we evaluate the average of $S(x; \Upsilon) S(x;
\Upsilon)$ and multiply the final result by $\rho(\bar x) \rho(\bar
x')$; this factor, however, disappears from $C_2$ [cf.\ prefactor in
Eq.~\eqref{eq:26}] and is shifted instead in the integration
\eqref{eq:24}.  Again, if $P_0$ is not vanishing, we need to correct
for cases where $S$ cannot be evaluated at $x$ or $x'$; The correcting
factor is still $1 / \bigl[ 1 - P_0(x, x') \bigr]$.

\subsection{Normalization}
\label{sec:normalization}

Almost every smoothing procedure used in Astronomy has a simple
normalization property: If all values are the same, i.e.\ $y_i = c$,
then the interpolating function $S$ always returns $c$ at every point
$x$:
\begin{equation}
  \label{eq:27}
  S \bigl( x; \bigl\{ (x_i, c) \bigr\} \bigr) = c \; .
\end{equation}
Normalized smoothing procedures are sometimes called ``unbiased.''\@
This property is satisfied, for example, by the interpolator
\eqref{eq:2}.  For a linear smoothing, this property immediately
implies
\begin{equation}
  \label{eq:28}
  \int_X K(x; \bar x) \rho(\bar x) \, \diff \bar x = 1 \qquad \forall
  x \in X\; .
\end{equation}
This can be easily verified by using a constant function $f(x) = 1$
and by noting that in this case $\bigl\langle \tilde f(x) \bigr\rangle
= 1$; Eq.~\eqref{eq:14} then gives the normalization of $K$.
Similarly, for the covariance kernels we have
\begin{align}
  \label{eq:29}
  & \int_X \diff \bar x \, \rho(\bar x) C_1(x, x'; \bar x) \notag\\
  & \quad {} + \int_X \diff \bar x \, \rho(\bar x) \int_X \diff \bar
  x' \, \rho(\bar x') C_2(x, x'; \bar x, \bar x') = 1 \; .
\end{align}
Note that the Poisson noise $T_\mathrm{P}$ vanishes if $f(x)$ is
``flat.''

Given any linear smoothing procedure $S$ written in the form
\eqref{eq:12}, we can obtain a related normalized interpolator $S'$:
\begin{align}
  \label{eq:30}
  S' \bigl( x; \bigl\{ (x_i, y_i) \bigr\} \bigr) = {} & \biggl[
  \sum_{j=1}^N
  w_j\bigl( x; \bigl\{ x_i \bigr\} \bigr) y_j \biggr] \notag\\
  & {} \times \biggl[ \sum_{j=1}^N w_j\bigl( x; \bigl\{ x_i \bigr\}
  \bigr) \biggr]^{-1} \; .
\end{align}
Indeed, in many cases interpolating techniques are directly written as
in Eq.~\eqref{eq:30}.  An example is given by our toy-interpolator
\eqref{eq:2}, which is of the form \eqref{eq:30} with
\begin{equation}
  \label{eq:31}
  w_j\bigl( x; \bigl\{ x_i \bigr\} \bigr) = 1 / |x - x_j|^\alpha \; .
\end{equation}

\subsection{Spatial symmetries}
\label{sec:spatial-symmetries}

Basically all interpolation procedures of some interest satisfy some
spatial symmetries.  In this section we will consider three common
spatial invariance properties, namely translation, rotation, and
scaling invariance.  These properties can have important consequences
on the forms of the three kernels $K$, $C_1$, and $C_2$, provided that
similar symmetries applies for the spatial distribution of location
measurements.

\subsubsection{Invariance upon translation}
\label{sec:invar-upon-transl}

Many smoothing techniques are invariant upon translation, i.e. there
is no ``preferred'' point on $X$ for the smoothing:
\begin{equation}
  \label{eq:32}
  S \bigl( x + d; \bigl\{ ( x_i + d, y_i) \bigr\} \bigr) = S \bigl( x;
  \bigl\{ (x_i, y_i) \bigr\} \bigr) \; .
\end{equation}
For example, the simple smoothing \eqref{eq:2} is invariant upon
translation because it depends only on the distances $| x - x_i|$
between the interpolated point $x$ and the locations $\{ x_i \}$.  A
linear smoothing invariant upon translation has necessarily associated
weights $w_i$ [cf.\ Eq.~\eqref{eq:12}] of the form $w_i = w_i \bigl(
\{ x_j - x \} \bigr)$.  Moreover, if the distribution of locations is
also invariant upon translation (this implies, among other things,
that the density $\rho$ is uniform), then the kernel $K$ is also
invariant upon translation, i.e.\ $K(x; \bar x) = K(x - \bar x)$, and
thus Eq.~\eqref{eq:14} becomes a simple convolution.

If the interpolation procedure is also normalized, than an interesting
property holds:
\begin{equation}
  \label{eq:33}
  \int_X \bigl\langle \tilde f(x) \bigr\rangle \, \diff x = \int_X
  f(x) \, \diff x \; .
\end{equation}
This can be shown by noting that for a smoothing procedure invariant
upon translation, $\bigl\langle \tilde f \bigr\rangle$ is the
convolution of $f$ with the kernel $K$.  Moreover, for a normalized
smoothing the kernel $K$ is normalized to unity [cf.\ 
Eq.~\eqref{eq:28}], and thus Eq.~\eqref{eq:33} holds.  Note that this
result basically states the conservation of the ``signal'': The total
measured signal is equal to the true, original signal.

The invariance upon translation has also consequences on the two
covariance kernels.  More specifically, they must be of the form
$C_1(x, x'; \bar x) = C_1(x - \bar x, x' - \bar x)$ and $C_2(x, x';
\bar x, \bar x') = C_2(x - \bar x, x' - \bar x; \bar x' - \bar x)$.
Note also that $P_0(x)$ is independent of $x$; hence, we will write
just $P_0$; similarly, $P_0(x, x')$ can depend only on $(x - x')$.

\subsubsection{Invariance upon rotation}
\label{sec:invar-upon-rotat}

Smoothing procedures in spaces $X$ of dimension larger than one are
also often invariant upon rotation and mirror symmetry, i.e. they are
isotropic.  Formally
\begin{equation}
  \label{eq:34}
  S \bigl( x; \bigl\{ \bigl( x + R (x_i - x), y_i \bigl) \bigr\}
  \bigr) = S \bigl( x; \bigl\{ (x_i, y_i) \bigr\} \bigr) \; .
\end{equation}
In this equation, $R$ is any orthogonal matrix; note that the rotation
is actually performed around the point $x$.  The smoothing
\eqref{eq:2} is invariant upon rotation (to show this we note, again,
that this interpolator only depends on the distances $|x - x_i|$).  

If this symmetry holds also for the generation of locations (this
implies, again, that $\rho$ is uniform) and if the interpolator is
linear, than the kernel $K$ is of the form $K(x; \bar x) = K'\bigl( x;
\lvert x - \bar x \rvert \bigr)$, i.e.\ the smoothing kernel can
depend on $x$ and on the distance $\lvert x - \bar x \rvert$ between
$x$ and $\bar x$ only.

The kernel associated with an interpolator invariant upon rotation has
an interesting property.  Let us evaluate the integral
\begin{align}
  \label{eq:35}
  \int_X (x - \bar x) K(x; \bar x) \, \diff \bar x = \int_X (x - \bar
  x) K\bigl( x; \lvert x - \bar x \rvert \bigr) \, \diff \bar x = 0 \;
  ,
\end{align}
where the last equality holds because the integrand is an odd function
of $(x - \bar x)$.  We can recast this result in a more interesting
form
\begin{equation}
  \label{eq:36}
  \biggl[ \int_X \bar x K(x; \bar x) \, \diff \bar x \biggr] \biggl[
  \int_X K(x; \bar x) \, \diff \bar x \biggr]^{-1} = x \; .
\end{equation}
This quantity in the left hand side of this equation is a measure of
the ``center of mass'' for the kernel $K$; hence Eq.~\eqref{eq:36}
basically assures that there is no systematic ``offset'' on the final
map.

If both spatial invariance properties considered so far hold for a
linear smoothing procedure, then the kernel $K$ associated with the
smoothing is of the form $K(x; \bar x) = K''\bigl( \lvert x - \bar x
\rvert \bigr)$, so that only distances are involved.  This property
also holds for the covariance kernels, which are only dependent on the
various distances between $x$, $x'$, $\bar x$, and (for $C_2$ alone)
$\bar x'$.  As a result, the noise term $T_\sigma$ is a function of
$\lvert x - x' \rvert$ alone.

\subsubsection{Invariance upon scaling}
\label{sec:invar-upon-scal}

A third spatial invariance occurs for some smoothing techniques, which
are intrinsically scale-free: The result of the smoothing depends only
on the \textit{ratios\/} of distances and not on their absolute
values.  Formally in terms of the smoothing function, we have
\begin{equation}
  \label{eq:37}
  S \bigl( x; \bigl\{ \bigl( x + k  (x_i - x), y_i \bigr) \bigr\}
  \bigr) = S \bigl( x; \bigl\{ (x_i , y_i) \bigr\} \bigr) \; ,
\end{equation}
where $k$ is an arbitrary positive real number.  It is not difficult
to verify that the interpolator \eqref{eq:2} is invariant upon
scaling.  Indeed, for any positive real $k$, we have
\begin{align}
  \label{eq:38}
  & S \bigl( x; \bigl\{ \bigl( x + k (x_i - x), y_i \bigr) \bigr\}
  \bigr) = \frac{\sum_{j=1}^N y_j / \bigl( k^\alpha | x - x_j |^\alpha
    \bigr)}{\sum_{j=1}^N 1 / \bigl( k^\alpha | x - x_j |^\alpha
    \bigr)} \notag\\
  & \qquad {} = \frac{\sum_{j=1}^N y_j / | x - x_j
    |^\alpha}{\sum_{j=1}^N 1 / | x - x_j |^\alpha} = S \bigl( x;
  \bigl\{ (x_i , y_i) \bigr\} \bigr) \; .
\end{align}

If measurement locations are uncorrelated and uniformly distributed
(i.e., for a homogeneous Poisson process), then the scale-free
property allows us to derive a number of consequences that greatly
simplify the analysis of the estimator.  For a scale-free linear
smoothing, the typical scale-length of the smoothing kernel $K$ is set
only by the density of objects $\rho$ (no other scales are available;
note instead that in presence of clustering we would immediately have
another scale, the correlation length).  Formally, this can be seen by
noting that two different location configurations, $\{ x_i \}$ and
$\bigl\{ x + k (x_i - x) \bigr\}$ have the same probability if the
density is changed from $\rho$ to $k^{-n} \rho$, where $n$ is the
space dimension.  As a result, all kernels are simply influenced by a
change of $\rho$:
\begin{align}
  \label{eq:39}
  & \rho \mapsto k^{-n} \rho \; , \\
  \label{eq:40}
  & K(\bar x + d; \bar x) \mapsto K(\bar x + kd; \bar x) \; , \\
  \label{eq:41}
  & C_1(\bar x + d, \bar x + d'; \bar x) \mapsto C_1(\bar x + k
  d, \bar x + k d'; \bar x) \; , \\
  \label{eq:42}
  & C_2(\bar x + d, \bar x + d'; \bar x, \bar x + \bar d) \mapsto
  \notag\\
  & \hskip 9em C_2(\bar x + k d, \bar x + k d'; \bar x,
  \bar x + k \bar d) \; .
\end{align}
Hence, we can study the kernels associated to a smoothing technique
invariant upon scaling for a given density (say $\rho = 1$) and then
generalize the results obtained using Eqs.~(\ref{eq:39}--\ref{eq:42}).
We also note that for a normalized smoothing, as $\rho \rightarrow
\infty$ the product $\rho(\bar x) K$ will be proportional to Dirac's
delta distribution.

\subsection{Exact interpolators}
\label{sec:exact-interp}

An interpolating procedure is said to be exact if it honors the data
points upon which the interpolation is based:
\begin{equation}
  \label{eq:43}
  S \bigl( x_j; \bigl\{ (x_i, y_i) \bigr\} \bigr) = y_j \; .
\end{equation}
Honoring data points is seen as an important feature in many
applications; on the other hand, if it is known that the data points
might be affected by significant errors, exact interpolators might not
be the right choice.

Regarding the exactness, the interpolator \eqref{eq:2} needs a few
words of explanation.  Indeed, when the interpolated position $x$
coincides with one of the locations $\{ x_i \}$, the expression
\eqref{eq:2} becomes singular, and thus the interpolator is
apparently undefined.  However, the limit $x \rightarrow x_i$ of $S$
is defined and equal to $y_i$ (we assume here that all locations are
different; the case of two or more coincident locations has no
statistical relevance).  Hence, we can safely extend the definition of
the smoothing technique \eqref{eq:2} to all points $x \in X$, with
the convention that limits must be taken if the expression is
undefined.  This way, we can state that our toy-interpolator is
exact.

Recalling the numerical method used to evaluate $K$ for a linear
interpolator, we can deduce that an exact estimator has the property
$K(\bar x; \bar x) = 1 / \bigl[ 1 - P_0(\bar x) \bigr]$.  In fact, if
we use the numerical method, we will always measure $S\bigl( \bar x;
\bigl\{ (x_i, y_i) \bigr\} \bigr) = 1$, since we have assigned a value
$1$ to the location $\bar x$, and thus the average of all those
measurements is also $1$ (here we are neglecting cases where the
smoothing is not defined).  Similarly, we find $C_1(\bar x, \bar x;
\bar x) = 1 / \bigl[ 1 - P_0(\bar x, \bar x) \bigr]$ and $C_2(\bar x,
\bar x'; \bar x, \bar x') = C_2(\bar x', \bar x; \bar x, \bar x') = 1
/ \bigl[ 1 - P_0(\bar x, \bar x') \bigr]$.

\subsection{Bounded interpolators}
\label{sec:bound-interp}

An interpolating procedure is said to be bounded if interpolated
values are always between the smallest and the largest measured
values:
\begin{equation}
  \label{eq:44}
  \min y_i \le S \bigl( x; \bigl\{ (x_i, y_i) \bigr\} \bigr) \le \max
  y_i \; .
\end{equation}
This is a rather natural property satisfied by many smoothing
techniques [e.g., the interpolator \eqref{eq:2}].  Note that a
bounded interpolator is always normalized, since Eq.~\eqref{eq:44}
implies Eq.~\eqref{eq:27}.

If a linear interpolator is bounded, then we can put superior and
inferior limits on $K(x; \bar x)$.  In fact, recalling again the
numerical technique used to evaluate $K$, we obtain
\begin{equation}
  \label{eq:45}
  0 \le K(x; \bar x) \le \bigl[ 1 - P_0(x) \bigr]^{-1} \; .
\end{equation}
Hence, the kernel $K$ is non-negative and $\bigl[ 1 - P_0(x)
\bigr]^{-1}$ is a superior limit for it.  Similarly, for the other
kernels we obtain
\begin{align}
  \label{eq:46}
  0 &\le C_1(x, x'; \bar x) \le \bigl[1 - P_0(x, x') \bigr]^{-1} \; ,
  \\
  \label{eq:47}
  0 &\le C_2(x, x'; \bar x, \bar x') \le \bigl[ 1 - P_0(x, x')
  \bigr]^{-1} \; .
\end{align}

We can go further in this analysis by defining the effective number of
objects used by the smoothing procedure as
\begin{align}
  \label{eq:48}
  & \mathcal{N}_\mathrm{eff}(x) = \frac{1}{1 - P_0(x)} \left[ \int_X
    K(x; \bar x) \rho(\bar x) \, \diff \bar x \right]^2 \notag\\
  & \quad {} \times \left[ \int_X \left[ K(x; \bar x) \right]^2
    \rho(\bar x) \, \diff \bar x \right]^{-1} \; .
\end{align}
Note that $\mathcal{N}_\mathrm{eff}$ can be defined for any linear
smoothing procedure; for a normalized (and in particular for a
bounded) interpolator, the first term in the right hand side of
Eq.~\eqref{eq:48} is unity.  For an interpolator invariant upon
translation, the quantity $\mathcal{N}_\mathrm{eff}(x)$ does not
depend on $x$, and thus we will write just $\mathcal{N}_\mathrm{eff}$.
Using the definition~\eqref{eq:48} we can recast the upper limit for
$K(x; \bar x)$, valid for a bounded interpolator, as a lower limit for
the effective number of objects $\mathcal{N}_\mathrm{eff}(x)$:
\begin{align}
  \label{eq:49}
  \mathcal{N}_\mathrm{eff}(x) & {} \ge \frac{1}{1 - P_0(x)} \left[
    \int_X \frac{1}{1 - P_0(x)} K(x; \bar x) \rho(\bar x) \, \diff
    \bar x \right]^{-1} \notag\\
  & {} \ge 1 \; .
\end{align}
Equation~\eqref{eq:49} put a lower limit to the
effective number for the kernel $K$, i.e.\ to the expected
\textit{resolution\/} of the smoothing.  Finally, we observe that
interpolators for which the scaling invariance holds have effective
number independent of $\rho$ (remember that $\rho$ is supposed to be
constant for scaling invariant interpolators).  

The covariance of a bounded interpolating procedure satisfies some
interesting properties.  Using $C_1 \ge 0$ [see \eqref{eq:46}]
in Eq.~\eqref{eq:29}, in fact, we obtain
\begin{equation}
  \label{eq:51}
  \int_X \diff \bar x \, \rho(\bar x) C_1(x, x'; \bar x) \le 1 \; ,
\end{equation}
so that $T_\sigma(x, x') \le \sigma^2$.  Hence, we have obtained an
upper limit for the contribution to the noise from measurement errors.
A comparison of this result with the inequality
$\mathcal{N}_\mathrm{eff} \ge 1$ suggests that our definition of
effective number of objects is sensible.

We can better appreciate the relationship between the covariance
kernel $C_1$ and $\mathcal{N}_\mathrm{eff}$ with the following
argument, valid for any linear interpolation procedure.  Let us
consider $C_1(x, x; \bar x)$ and briefly recall how this quantity can
be measured using the numerical approach discussed in
Sect.~\ref{sec:covariance-linear}.  We generate locations, keeping one
object with weight $1$ in $\bar x$, and take the average of the
quantity $S(x; \Upsilon_1) S(x; \Upsilon_1)$; finally we multiply the
average by $1 / \bigl[ 1 - P_0(x) \bigr]$, because in this particular
case $P_0(x, x) = P_0(x)$.  Something quite similar is done for $K(x;
\bar x)$ (see Sect.~\ref{sec:bias-linear}): The configurations used
are the same, in the sense that $\Upsilon = \Upsilon_1$, but in this
case we take the average of $S(x; \Upsilon)$.  From this discussion,
we see that $K(x; \bar x)$ is the simple average of a set of some
non-negative numbers, while $C_1(x, x; \bar x)$ is, apart from a known
numerical factor, the average of the squares of the same quantities.
Hence, the following inequality holds:
\begin{equation}
  \label{eq:52}
  C_1(x, x; \bar x) \ge \bigl[ 1 - P_0(x) \bigr] \bigl[ K(x;
  \bar x) \bigr]^2 \; .
\end{equation}
For a normalized interpolator, this equation implies $T_\sigma(x,x)
\ge \sigma^2 / \mathcal{N}_\mathrm{eff}$.  Hence, the effective number
of objects puts a lower limit on the measurement error.

\subsection{Local interpolators}
\label{sec:local-interpolators}

An interpolator is called local if the interpolated value at $x$
depends only on the values $y_i$ for locations $x_i$ ``close'' to $x$;
an interpolator which is not local is said to be global.  Global
interpolators usually produce smoother maps but often require more
computing time.  The interpolator \eqref{eq:2} is global.

We further distinguish between strongly and weakly local interpolators
depending on the meaning that is given to the word ``close'' in the
previous definition.  An interpolator is strongly local if it uses
only values $y_i$ whose locations $x_i$ fall within a fixed range from
$x$; a weakly local interpolator, instead, is only guaranteed to use a
\textit{finite\/} number of points close to $x$.  We stress that this
definition is not standard but is needed for a rigorous
characterization of local smoothing techniques.  Note that a strongly
local interpolator cannot be scale invariant, because it depends only
on points within a \textit{fixed range\/} and hence, there is a
natural scale for the interpolator, its range.  A strongly local
interpolator has always $P_0(x) > 0$, because the interpolator cannot
be defined if there is a large ``void'' without points around $x$.

Using again the numerical technique to obtain $K$, we immediately find
that the kernel $K$ associated with a strongly local interpolator has
compact support, i.e.\ $K(x; \bar x)$ vanishes if $\lvert x - \bar x
\rvert$ is large; analogously, $C_1(x, x'; \bar x)$ vanishes if either
$\lvert x - \bar x \rvert$ or $\lvert x' - \bar x \rvert$ are large,
and $C_2(x, x'; \bar x, \bar x')$ vanishes if $x$ or $x'$ are far away
from $\bar x$ or $\bar x'$.  Note also that for strong local
estimators, if $\lvert \bar x - \bar x' \rvert$ is large, $C_2(x, x';
\bar x, \bar x')$ converges to
\begin{align}
  \label{eq:53}
  & C_2(x, x'; \bar x, \bar x') = \frac{\bigl[ 1 - P_0(x) \bigr]
    \bigl[ 1 - P_0(x') \bigr]}{\bigl[ 1 - P_0(x, x') \bigr]} \notag\\
  & \qquad {} \times \bigl[ K(x; \bar x) K(x'; \bar x') + K(x'; \bar
  x) K(x; \bar x') \bigr] \; .
\end{align}
Not much, instead, can be said regarding weakly local interpolators;
indeed, weakly local interpolators have been defined here mainly to
have a complete classification of locality.

Several strongly local interpolators can be defined with just a single
point inside their range.  In this case, we can show that in the limit
$\rho \rightarrow 0$, the kernel $K$ associated to a interpolator
approaches a top-hat function, provided the density $\rho$ is uniform.
For if $\rho$ is sufficiently small everywhere, we expect a negligible
probability, when using the numerical technique to obtain $K(x; \bar
x)$, of having points other than $\bar x$ inside the range of the
function.  In this case, all terms in the sums of Eq.~\eqref{eq:30}
vanish except for $\bar x$.  As a result, we always measure $1$ for
points inside the range of the interpolator, and $0$ outside.
Moreover, if $\pi$ is set around the point $x$ where the interpolator
is defined, we find [cf.\ Eq.~\eqref{eq:69}]
\begin{equation}
  \label{eq:54}
  P_0(x) = \exp \biggl( -\int_\pi \rho(x') \, \diff x' \biggr) 
  \simeq 1 - \int_\pi \rho(x') \, \diff x' \; .
\end{equation}
Notice that the dependence on $x$ of $K(x; \bar x)$ is also expressed
by the set $\pi$, which is a function of the location $x$.  In
summary, we obtain
\begin{equation}
  \label{eq:55}
  K(x; \bar x) \simeq
  \begin{cases}
    1 / \int_\pi \rho(x') \, \diff x' &
    \text{if $\bar x \in \pi \; ,$} \\
    0 & \text{otherwise$\; .$}
  \end{cases}
\end{equation}
As expected, this kernel is manifestly normalized according to
Eq.~\eqref{eq:27}.  In case of a uniform distribution $\rho$, this
expression simplifies to
\begin{equation}
  \label{eq:56}
  K(x; \bar x) \simeq
  \begin{cases}
    1 / \rho \mu(\pi) &
    \text{if $\bar x \in \pi \; ,$} \\
    0 & \text{otherwise$\; ,$}
  \end{cases}
\end{equation}
where $\mu(\pi)$ is the area of the set $\pi$.  Hence, in the limit
$\rho \rightarrow 0$ we reach the upper bound for $K$ stated in
Eq.~\eqref{eq:45}.  Regarding the covariance kernel, calling $\mu(\pi
\cap \pi')$ the area of the intersection of the ranges for $x$ and
$x'$, we obtain (here, for simplicity, we assume a uniform location
density $\rho$)
\begin{align}
  \label{eq:57}
  & C_1(x, x'; \bar x) \simeq
  \begin{cases}
    1/ \rho \mu(\pi \cap \pi) & \text{if $\bar x \in \pi \cap \pi \;
    ,$} \\
    0 & \text{otherwise$\; ,$}
  \end{cases}\\
  \label{eq:58}
  & C_2(x, x'; \bar x, \bar x') \simeq 0 \; .
\end{align}
Note that $C_2$ vanishes in the limit $\rho \rightarrow 0$ (this is
due to the extra factor $\rho$, with respect to $C_1$, present in the
evaluation of $C_2$).  Note also that in this case we have reached the
limit $T_\sigma = \sigma^2$ discussed after Eq.~\eqref{eq:51}.
Finally, we can verify without difficulties that the normalization of
Eq.~\eqref{eq:29} is satisfied.

\subsection{Other properties}
\label{sec:other-properties}

Recalling again the numerical technique to obtain $K$ described in
Sect.~\ref{sec:bias-linear}, we can consider a number of other
classifications all focused on properties of $S$ when all but one
point are vanishing.

For example, suppose that the smoothed map $S$ for the configuration
$\Upsilon = \bigl\{ (\bar x, 1) \bigr\} \cup \bigl\{ (x_i, 0) \bigr\}$
is smaller than unity at each point.  Such a smoothing procedure could
not be bounded, but we can still obtain the upper limit $K(x; \bar x)
< 1 / \bigl( 1 - P_0(x) \bigr)$ for $K(x; \bar x)$ (as in case of
bounded interpolators) and Eq.~\eqref{eq:49} holds.

Another interesting property to consider is monotonicity.  Suppose
that, for the same configuration $\Upsilon$ considered above, the
function $S$ is monotonically decreasing as $\lvert x - \bar x \rvert$
increases: Then $K(x; \bar x)$ will also decrease monotonically with
$\lvert x - \bar x \rvert$.  If the interpolator is exact, then, this
property implies $K(x; \bar x) \le 1 / \bigl[ 1 - P_0(x) \bigr]$,
because $K(x, x) = 1 / \bigl[ 1 - P_0(x) \bigr]$, and again we recover
Eq.~\eqref{eq:49}.

\section{Kernels of some smoothing procedures}
\label{sec:kern-some-smooth}

In this section we will consider several smoothing procedures and
derive, analytically or numerically, the relative kernels $K$; for
simplicity, we will not consider the kernels $C_1$ and $C_2$;
moreover, we will assume a uniform spatial distribution of locations
characterized by a density $\rho$ and no correlation.  A standard
approach will be followed for each interpolator.  We will first
briefly introduce and define the interpolator; then we will classify
it according to the nomenclature introduced in
Sect.~\ref{sec:classifications}; finally we will evaluate the kernel
$K$.  

The interpolation techniques discussed below are generally well known
and are described in any book on spatial interpolation
\citep[e.g.][]{Watson}.  As a reference (extremely useful also for a
deep discussion on statistics of spatial data) we refer to
\citet{Cressie}, and in particular to Sect.~5.9 of this book
\citep[see also][]{Adler,Preston}.

\subsection{Nearest neighbor}
\label{sec:nearest-neighbor}

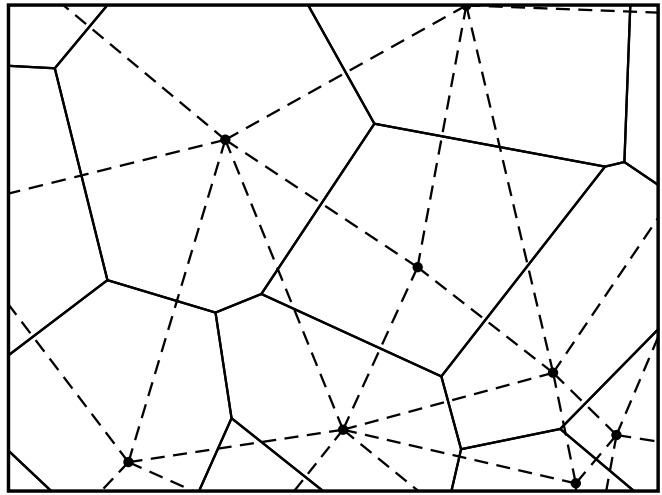
\begin{figure}[!t]
  \begin{center}
    \resizebox{\hsize}{!}{%
      \begin{pspicture}(-4, -3)(4, 3)
  \psframe[linewidth=1.2pt](-4, -3)(4, 3)
  \begin{psclip}{\psframe[linestyle=none](-4, -3)(4, 3)}
    \psset{dotstyle=*}
    \psdots(-6.54279, 6.64948)
    \psdots(3.60819, -5.29611)
    \psdots(8.34156, -1.34826)
    \psdots(8.35021, -0.926710)
    \psdots(5.33558, 4.77643)
    \psdots(2.97002, -2.88506)
    \psdots(-3.31578, 8.66421)
    \psdots(0.119059, -2.22682)
    \psdots(3.04365, -4.60810)
    \psdots(-6.83653, 5.92650)
    \psdots(8.25503, 1.90329)
    \psdots(-4.84814, 0.452290)
    \psdots(6.21980, -2.72900)
    \psdots(-4.65384, 4.09800)
    \psdots(-6.22256, -9.23018)
    \psdots(-5.25355, 9.10374)
    \psdots(-3.75470, -3.94357)
    \psdots(1.03208, -0.234371)
    \psdots(8.89767, -5.38331)
    \psdots(3.46929, -2.29182)
    \psdots(2.26604, 6.15143)
    \psdots(-8.25140, -7.21827)
    \psdots(5.64105, 2.82342)
    \psdots(-2.50932, -2.62509)
    \psdots(-8.40006, -1.15447)
    \psdots(1.62920, 2.97474)
    \psdots(-1.32273, 1.32737)
    \psdots(-0.803511, -3.38555)
    \psdots(2.69288, -1.52929)
    \psdots(-6.35886, 1.38174)
    \psline(-11.1801, 8.11242)(-5.18767, 5.67775)(-4.16593, 6.43418)(-4.66061, 7.22651)(-11.1801, 10.6513)(-11.1801, 8.11242)
    \psline(-0.613698, -8.99608)(5.92336, -25.3314)(6.25236, -5.37412)(4.71620, -3.81133)(-0.0686409, -7.73748)(-0.613698, -8.99608)
    \psline(8.18345, -3.42590)(22.6586, -1.43097)(6.67192, -1.10316)(8.18345, -3.42590)
    \psline(6.26387, 0.419724)(5.71750, 0.0250157)(6.67192, -1.10316)(22.6586, -1.43097)(49.5851, 1.87673)(6.26387, 0.419724)
    \psline(3.27782, 4.29641)(3.65830, 3.51370)(7.56823, 4.12524)
    \psline(1.56531, -2.46599)(1.25233, -3.82156)(4.51900, -3.68196)(2.77930, -2.21784)(1.56531, -2.46599)
    \psline(-4.66061, 7.22651)(-4.16593, 6.43418)(-1.33643, 5.60504)
    \psline(-1.44288, -0.790904)(-1.24662, -2.08611)(1.12676, -3.97578)(1.25233, -3.82156)(1.56531, -2.46599)(1.32337, -1.57327)(-0.881190, -0.563049)(-1.44288, -0.790904)
    \psline(1.12676, -3.97578)(-0.0686409, -7.73748)(4.71620, -3.81133)(4.65438, -3.70684)(4.51900, -3.68196)(1.25233, -3.82156)(1.12676, -3.97578)
    \psline(-11.1801, 3.17249)(-11.1801, 3.17249)(-6.91046, 3.62125)(-5.18767, 5.67775)(-11.1801, 8.11242)(-11.1801, 3.17249)
    \psline(7.56823, 4.12524)(6.26387, 0.419724)(49.5851, 1.87673)
    \psline(-6.52224, -0.576278)(-5.82214, -2.12391)(-4.66520, -1.83613)(-2.76691, -0.393413)(-3.41158, 2.20376)(-4.76739, 2.27602)(-6.52224, -0.576278)
    \psline(4.65438, -3.70684)(4.71620, -3.81133)(6.25236, -5.37412)(8.18345, -3.42590)(6.67192, -1.10316)(5.71750, 0.0250157)(5.23963, -0.0247954)(4.65438, -3.70684)
    \psline(-5.18767, 5.67775)(-6.91046, 3.62125)(-4.76739, 2.27602)(-3.41158, 2.20376)(-1.29579, 4.74756)(-1.21540, 5.19719)(-1.33643, 5.60504)(-4.16593, 6.43418)(-5.18767, 5.67775)
    \psline(5.92336, -25.3314)(5.92336, -25.3314)(-0.613698, -8.99608)(-1.41074, -8.25708)(-5.41563, -6.38755)(-11.1801, -12.2006)(-11.1801, -25.3314)(5.92336, -25.3314)
    \psline(-11.1801, 10.6513)(-4.66061, 7.22651)(-11.1801, 10.6513)
    \psline(-5.82214, -2.12391)(-7.04155, -4.15489)(-5.41563, -6.38755)(-1.41074, -8.25708)(-2.18118, -4.18245)(-4.66520, -1.83613)(-5.82214, -2.12391)
    \psline(-0.881190, -0.563049)(1.32337, -1.57327)(3.32852, 0.998442)(0.503078, 1.52417)(-0.881190, -0.563049)
    \psline(49.5851, 1.87673)(22.6586, -1.43097)(8.18345, -3.42590)(6.25236, -5.37412)(5.92336, -25.3314)(5.92336, -25.3314)
    \psline(2.77930, -2.21784)(4.51900, -3.68196)(4.65438, -3.70684)(5.23963, -0.0247954)(5.02333, 0.0670380)(2.77930, -2.21784)
    \psline(-1.33643, 5.60504)(-1.21540, 5.19719)(3.27782, 4.29641)(-1.33643, 5.60504)
    \psline(-11.1801, -12.2006)(-5.41563, -6.38755)(-7.04155, -4.15489)(-11.1801, -4.25635)(-11.1801, -12.2006)
    \psline(3.65830, 3.51370)(3.56555, 1.05442)(5.02333, 0.0670380)(5.23963, -0.0247954)(5.71750, 0.0250157)(6.26387, 0.419724)(7.56823, 4.12524)(3.65830, 3.51370)
    \psline(-4.66520, -1.83613)(-2.18118, -4.18245)(-1.24662, -2.08611)(-1.44288, -0.790904)(-2.76691, -0.393413)(-4.66520, -1.83613)
    \psline(-11.1801, -4.25635)(-7.04155, -4.15489)(-5.82214, -2.12391)(-6.52224, -0.576278)(-11.1801, 3.17249)(-11.1801, 3.17249)(-11.1801, -4.25635)
    \psline(-1.29579, 4.74756)(0.503078, 1.52417)(3.32852, 0.998442)(3.56555, 1.05442)(3.65830, 3.51370)(3.27782, 4.29641)(-1.21540, 5.19719)(-1.29579, 4.74756)
    \psline(-3.41158, 2.20376)(-2.76691, -0.393413)(-1.44288, -0.790904)(-0.881190, -0.563049)(0.503078, 1.52417)(-1.29579, 4.74756)(-3.41158, 2.20376)
    \psline(-1.24662, -2.08611)(-2.18118, -4.18245)(-1.41074, -8.25708)(-0.613698, -8.99608)(-0.0686409, -7.73748)(1.12676, -3.97578)(-1.24662, -2.08611)
    \psline(3.32852, 0.998442)(1.32337, -1.57327)(1.56531, -2.46599)(2.77930, -2.21784)(5.02333, 0.0670380)(3.56555, 1.05442)(3.32852, 0.998442)
    \psline(-11.1801, 3.17249)(-6.52224, -0.576278)(-4.76739, 2.27602)(-6.91046, 3.62125)(-11.1801, 3.17249)
    \psset{linestyle=dashed}
    \psline(8.35021, -0.926710)(8.34156, -1.34826)
    \psline(-3.31578, 8.66421)(-6.54279, 6.64948)
    \psline(-3.31578, 8.66421)(5.33558, 4.77643)
    \psline(0.119059, -2.22682)(2.97002, -2.88506)
    \psline(3.04365, -4.60810)(0.119059, -2.22682)
    \psline(3.04365, -4.60810)(3.60819, -5.29611)
    \psline(3.04365, -4.60810)(2.97002, -2.88506)
    \psline(-6.83653, 5.92650)(-6.54279, 6.64948)
    \psline(8.25503, 1.90329)(5.33558, 4.77643)
    \psline(8.25503, 1.90329)(8.35021, -0.926710)
    \psline(6.21980, -2.72900)(3.04365, -4.60810)
    \psline(6.21980, -2.72900)(3.60819, -5.29611)
    \psline(6.21980, -2.72900)(8.34156, -1.34826)
    \psline(6.21980, -2.72900)(8.35021, -0.926710)
    \psline(-4.65384, 4.09800)(-6.54279, 6.64948)
    \psline(-4.65384, 4.09800)(-6.83653, 5.92650)
    \psline(-4.65384, 4.09800)(-4.84814, 0.452290)
    \psline(-4.65384, 4.09800)(-3.31578, 8.66421)
    \psline(-6.22256, -9.23018)(3.60819, -5.29611)
    \psline(-5.25355, 9.10374)(-6.54279, 6.64948)
    \psline(-5.25355, 9.10374)(-3.31578, 8.66421)
    \psline(-3.75470, -3.94357)(-4.84814, 0.452290)
    \psline(-3.75470, -3.94357)(-6.22256, -9.23018)
    \psline(1.03208, -0.234371)(0.119059, -2.22682)
    \psline(8.89767, -5.38331)(8.25503, 1.90329)
    \psline(8.89767, -5.38331)(8.35021, -0.926710)
    \psline(8.89767, -5.38331)(8.34156, -1.34826)
    \psline(8.89767, -5.38331)(6.21980, -2.72900)
    \psline(8.89767, -5.38331)(3.60819, -5.29611)
    \psline(8.89767, -5.38331)(-6.22256, -9.23018)
    \psline(3.46929, -2.29182)(2.97002, -2.88506)
    \psline(3.46929, -2.29182)(3.04365, -4.60810)
    \psline(3.46929, -2.29182)(6.21980, -2.72900)
    \psline(2.26604, 6.15143)(-3.31578, 8.66421)
    \psline(2.26604, 6.15143)(-4.65384, 4.09800)
    \psline(2.26604, 6.15143)(5.33558, 4.77643)
    \psline(-8.25140, -7.21827)(-6.22256, -9.23018)
    \psline(-8.25140, -7.21827)(-3.75470, -3.94357)
    \psline(5.64105, 2.82342)(5.33558, 4.77643)
    \psline(5.64105, 2.82342)(3.46929, -2.29182)
    \psline(5.64105, 2.82342)(6.21980, -2.72900)
    \psline(5.64105, 2.82342)(8.35021, -0.926710)
    \psline(5.64105, 2.82342)(8.25503, 1.90329)
    \psline(-2.50932, -2.62509)(-4.84814, 0.452290)
    \psline(-2.50932, -2.62509)(-3.75470, -3.94357)
    \psline(-2.50932, -2.62509)(0.119059, -2.22682)
    \psline(-8.40006, -1.15447)(-8.25140, -7.21827)
    \psline(-8.40006, -1.15447)(-3.75470, -3.94357)
    \psline(-8.40006, -1.15447)(-4.84814, 0.452290)
    \psline(-8.40006, -1.15447)(-6.83653, 5.92650)
    \psline(1.62920, 2.97474)(-4.65384, 4.09800)
    \psline(1.62920, 2.97474)(1.03208, -0.234371)
    \psline(1.62920, 2.97474)(5.64105, 2.82342)
    \psline(1.62920, 2.97474)(5.33558, 4.77643)
    \psline(1.62920, 2.97474)(2.26604, 6.15143)
    \psline(-1.32273, 1.32737)(-4.65384, 4.09800)
    \psline(-1.32273, 1.32737)(-4.84814, 0.452290)
    \psline(-1.32273, 1.32737)(-2.50932, -2.62509)
    \psline(-1.32273, 1.32737)(0.119059, -2.22682)
    \psline(-1.32273, 1.32737)(1.03208, -0.234371)
    \psline(-1.32273, 1.32737)(1.62920, 2.97474)
    \psline(-0.803511, -3.38555)(0.119059, -2.22682)
    \psline(-0.803511, -3.38555)(-2.50932, -2.62509)
    \psline(-0.803511, -3.38555)(-3.75470, -3.94357)
    \psline(-0.803511, -3.38555)(-6.22256, -9.23018)
    \psline(-0.803511, -3.38555)(3.60819, -5.29611)
    \psline(-0.803511, -3.38555)(3.04365, -4.60810)
    \psline(2.69288, -1.52929)(1.62920, 2.97474)
    \psline(2.69288, -1.52929)(1.03208, -0.234371)
    \psline(2.69288, -1.52929)(0.119059, -2.22682)
    \psline(2.69288, -1.52929)(2.97002, -2.88506)
    \psline(2.69288, -1.52929)(3.46929, -2.29182)
    \psline(2.69288, -1.52929)(5.64105, 2.82342)
    \psline(-6.35886, 1.38174)(-6.83653, 5.92650)
    \psline(-6.35886, 1.38174)(-8.40006, -1.15447)
    \psline(-6.35886, 1.38174)(-4.84814, 0.452290)
    \psline(-6.35886, 1.38174)(-4.65384, 4.09800)
  \end{psclip}
\end{pspicture}}
    \caption{The relationship between Voronoi cells (solid lines) and Delaunay
      triangles (dashed lines).  The figure has been made by
      calculating the Delaunay triangulation and Voronoi tessellation
      for a large set of points on the plane, and by showing a small
      region.  Note that two points are connected in the Delaunay
      triangulation if and only if their Voronoi cells have a side in
      common.}
    \label{fig:1}
  \end{center}
\end{figure}

This interpolator, called also ``proximal,'' is probably the simplest
one can imagine: The value in a point is assumed to be equal to the
value of the nearest known point.  Hence, the output data for such
interpolator consists in Voronoi cells \citep[see][]{Okabe} centered
on the points $\{ x_i \}$ with abrupt changes at the boundaries (see
Fig.~\ref{fig:1}).

This interpolator is manifestly linear, normalized, invariant upon
translation, rotation, and scale, exact, bounded, and weakly local.
As a result, we expect a kernel $K$ of the form $K(x; \bar x) = K
\bigl( \lvert x - \bar x \rvert \bigr)$ which scales with the density
according to Eqs.~\eqref{eq:39} and \eqref{eq:40}.  Moreover, we
expect $P_0(x) = P_0(x, x') = 0$, $\mathcal{N}_\mathrm{eff}$
independent of $\rho$, $K(0) = 1$, and $K(r) \le 1$.  Finally, $K(r)$
will be not increasing.

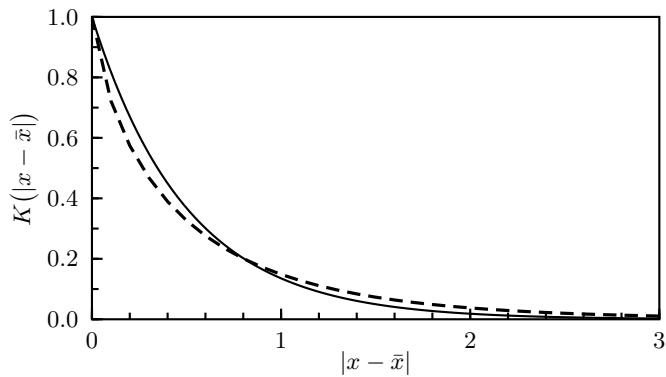
\begin{figure}[!t]
  \begin{center}
    \savedata{\figv}[%
    0.0 1.000000
    0.1 0.722545
    0.2 0.574201
    0.3 0.469115
    0.4 0.389368
    0.5 0.326644
    0.6 0.276184
    0.7 0.234947
    0.8 0.200852
    0.9 0.172404
    1.0 0.148496
    1.1 0.128281
    1.2 0.111104
    1.3 0.0964455
    1.4 0.0838899
    1.5 0.0731008
    1.6 0.0638032
    1.7 0.0557706
    1.8 0.0488153
    1.9 0.0427803
    2.0 0.0375343
    2.1 0.0329663
    2.2 0.0289827
    2.3 0.0255036
    2.4 0.0224613
    2.5 0.0197977
    2.6 0.0174630
    2.7 0.0154145
    2.8 0.0136152
    2.9 0.0120336
    3.0 0.0106419]
    \psset{xunit=2.5cm, yunit=4cm}
    \resizebox{\hsize}{!}{%
      \begin{pspicture}(-0.45,-0.15)(3.05,1.03)
        \psset{dimen=middle}
        \small
        \begin{psclip}{\psframe[linestyle=none](0,0)(3,1)}
          \dataplot[linewidth=1.2pt, linestyle=dashed]{\figv}
          \psplot[plotpoints=100]{0}{3}{2.71828182849 -2 x mul exp}
        \end{psclip}
        \psframe(0,0)(3,1)
        \multido{\n=0.0+0.2}{16}{%
          \psline(\n, 0)(\n, 2.5pt)
          }
        \multido{\n=0+1}{4}{%
          \psline(\n, 0)(\n, 4pt)
          \uput[-90](\n,0){$\n$}
          }
        \multido{\n=0.0+0.1}{11}{%
          \psline(0, \n)(2.5pt, \n)
          }
        \multido{\n=0.0+0.2}{6}{%
          \psline(0, \n)(4pt, \n)
          \uput[180](0,\n){$\n$}
          }
        \uput{12pt}[-90](1.5,0){$\lvert x - \bar x \rvert$}
        \uput{20pt}[180]{90}(0,0.5){$K\bigl( \lvert  x - \bar x
          \rvert \bigr)$}
      \end{pspicture}}
    \caption{The kernels $K$ associated with the one-dimensional
      nearest neighbor (solid line) and Delaunay interpolators (dashed
      line) for $\rho = 1$.  The kernel $K$ for the natural neighbor
      smoothing is identical to the one for the Delaunay interpolator.
      Since the interpolators considered are invariant upon
      translation and scaling, the kernels associated to different
      densities can be evaluated using the transformation
      \eqref{eq:39} and \eqref{eq:40}.  Note also that for both plots
      we find $K(0) = 1$ and $K(x; \bar x) \le 1$ because the
      smoothing techniques are exact and bounded.}
    \label{fig:2}
  \end{center}
\end{figure}

For this estimator, we can actually carry out calculations
analytically using the approach discussed in
Sect.~\ref{sec:bias-linear}.  Let us consider a point $x$ at distance
$r$ from $\bar x$.  The probability that the closest location for $x$
be $\bar x$ is given by $p(r) = \exp (- \rho k_n r^n)$, where $n$ is
the dimension of $X$ and $k_n$ is the \textit{measure} (area or
volume) of the unit sphere (e.g., $k_1 = 2$, $k_2 = \pi$, and $k_3 = 4
\pi / 3$).  In fact, $\bar x$ is the closest location to $x$ if the
ball of radius $r$ centered on $x$ does not contain any other
location.  The number of locations inside this ball follows a Poisson
distribution with average $\rho k_n r^n$, and thus this number
vanishes with the probability $p(r)$ given above.  Hence, the point at
$x$ will have a value $1$ with probability $p(r)$, and a vanishing
value with probability $1 - p(r)$.  In summary we obtain
\begin{equation}
  \label{eq:59}
  K\bigl( \lvert  x - \bar x \rvert \bigr) = \exp\bigl(- \rho k_n
  \lvert x - \bar x \rvert^n \bigr) \; .
\end{equation}
Using this expression for $K$ we can check all the properties stated
in Sect.~\ref{sec:classifications} (cf.\ also Figs.~\ref{fig:2} and
\ref{fig:3}).  Finally, we have $\mathcal{N}_\mathrm{eff} = 2$.

\subsection{Delaunay triangulation}
\label{sec:dela-triang}

The nearest neighbor technique is very simple but unfortunately
produces abrupt changes at the boundaries of the Voronoi cells, i.e.\ 
this interpolator is not smooth.  This problem can be solved by using
$n+1$ near points (where $n$ is the space dimension) and by linearly
interpolating their values.  For example, if $X = \R$, so that $n =
1$, we take the point at the left and the point at the right of $x$
and then use linear interpolation.  If $n > 1$, however, we have the
problem of choosing the most convenient points to perform the linear
interpolation (mathematically speaking, $\R^n$ is not totally ordered
for $n > 1$).  A standard approach is to use Delaunay triangulation
\citep[see, e.g.][]{Delaunay}, that we describe here for $n = 2$.  A
Delaunay triangulation for the locations $\{ x_i \}$ is the unique
triangulation of the plane with the property that no point in the set
$\{ x_i \}$ falls in the interior of the circumcircle (circle that
passes through all three vertices) of any triangle in the
triangulation.  This definition is easily generalized for $n > 2$.
The Delaunay triangulation for the points $\{ x_i \}$ is also closely
related to the Voronoi cells for the same points, one being the dual
of the other.  For example, the Delaunay triangulation can be obtained
by connecting all points whose Voronoi cells have a side in common
(see Fig.~\ref{fig:1}).

The Delaunay triangulation interpolator is thus obtained in the
following way.  The Delaunay triangulation is calculated for the
points $\{ x_i \}$; the interpolated value at $x$ is then obtained
using linear interpolation of the values for the three points that are
the vertices of the triangle to which $x$ belongs.  Delaunay
interpolation, or variants of it, has found already many applications
in the astrophysical context \citep[see, e.g.][]{1994A&A...283..361V,
  1996MNRAS.279..693B, 2000A&A...363L..29S}.

The Delaunay triangulation interpolator is linear, normalized,
invariant upon translations, rotations, and scaling, exact, bounded,
and weakly local.  We expect for $K$ the same properties as for the
nearest neighbor.

Analytical calculations for the Delaunay triangulation linear
smoothing are trivial only for dimension $n = 1$.  In this case, in
fact, we can just use the point at the left and the point at the right
of $x$ and proceed with linear interpolation.  Calling $d_- \ge 0$ and
$d_+ \ge 0$ the distances between the closest locations at the left
and at the right of $x$ and $x$ itself, we have
\begin{equation}
  \label{eq:60}
  \tilde f(x) = \frac{d_- f(x + d_+) + d_+ f(x - d_-)}{d_+ + d_-} \; .
\end{equation}
The probability distribution for both $d_+$ and $d_-$ is
\begin{equation}
  \label{eq:61}
  p(d_\pm) = \rho \e^{-\rho d_{\pm}} \; ,
\end{equation}
and thus we obtain
\begin{align}
  \label{eq:62}
  \bigl\langle \tilde f(x) \bigr\rangle = {} & \rho^2 \int_0^\infty
  \e^{-\rho d_+} \diff d_+ \notag\\
  & {} \times \int_0^\infty \e^{-\rho d_-} \frac{f(x + d_+) d_- + f(x
    - d_-) d_+}{d_+ + d_-} \, \diff d_- \notag\\
  {} = {} & \rho^2 \int_{-\infty}^\infty f(x + x_1) \e^{-\rho \lvert
    x_1 \rvert} \, \diff x_1 \int_0^\infty \frac{x_2 \e^{-\rho
      x_2}}{\lvert x_1 \rvert + x_2} \,
  \diff x_2 \notag\\
  {} = {} & \rho \int_{-\infty}^\infty f(\bar x) K(x - \bar x) \,
  \diff \bar x \; ,
\end{align}
where the kernel $K$ is given by
\begin{equation}
  \label{eq:63}
  K(x) = \e^{-\rho \lvert x \rvert} - \rho \lvert x \rvert
  \Gamma(0, \rho \lvert x \rvert) \; .
\end{equation}
Here $\Gamma$ is the incomplete gamma function, defined as
\begin{equation}
  \label{eq:64}
  \Gamma(a, x) = \int_x^\infty t^{a - 1} \e^{-t} \, \diff t \; .
\end{equation}
The same result is obtained using the numerical technique of
Sect.~\ref{sec:bias-linear}.  It can be verified that, as expected,
$K(0) = 1$ and $K(x) \le 1$; moreover, $K$ depends only on the product
$\rho \lvert x \rvert$, so that the scaling invariance
(\ref{eq:39}--\ref{eq:40}) holds.  For this kernel we find
$\mathcal{N}_\mathrm{eff} \simeq 2.44417$.  A plot of $K(x)$ for $n =
1$ is shown in Fig.~\ref{fig:2}.

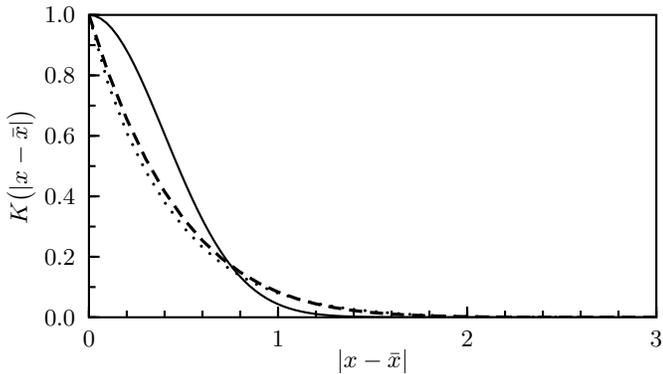
\begin{figure}[!t]
  \begin{center}
    \input fig3.tex
    \input fig4.tex
    \psset{xunit=2.5cm, yunit=4cm}
    \resizebox{\hsize}{!}{%
      \begin{pspicture}(-0.45,-0.15)(3.05,1.03)
        \psset{dimen=middle}
        \small
        \begin{psclip}{\psframe[linestyle=none](0,0)(3,1)}
          \dataplot[linewidth=1.2pt, linestyle=dotted]{\figiii}
          \dataplot[linewidth=1.2pt, linestyle=dashed]{\figiv}
          \psplot[plotpoints=100]{0}{3}{2.71828182849 -3.141592654 x
          dup mul mul exp}
        \end{psclip}
        \psframe(0,0)(3,1)
        \multido{\n=0.0+0.2}{16}{%
          \psline(\n, 0)(\n, 2.5pt)
          }
        \multido{\n=0+1}{4}{%
          \psline(\n, 0)(\n, 4pt)
          \uput[-90](\n,0){$\n$}
          }
        \multido{\n=0.0+0.1}{11}{%
          \psline(0, \n)(2.5pt, \n)
          }
        \multido{\n=0.0+0.2}{6}{%
          \psline(0, \n)(4pt, \n)
          \uput[180](0,\n){$\n$}
          }
        \uput{12pt}[-90](1.5,0){$\lvert x - \bar x \rvert$}
        \uput{20pt}[180]{90}(0,0.5){$K\bigl( \lvert  x - \bar x
          \rvert \bigr)$}
      \end{pspicture}}
    \caption{The kernels $K$ associated with the two-dimensional
      nearest neighbor (solid line), Delaunay triangulation (dashed
      line), and natural neighbor (dotted line) interpolators.  For
      all plots we have used a density $\rho = 1$.  Since all
      smoothing techniques are invariant upon translation, rotation,
      and scaling, the kernels associated to different densities can
      be evaluated using the transformation \eqref{eq:39} and
      \eqref{eq:40}.}
    \label{fig:3}
  \end{center}
\end{figure}

For $n > 1$ calculations cannot be easily carried out analytically.
However, since the smoothing procedure is expected to be scale
invariant, we can numerically evaluate $K$ for, say, $\rho = 1$, and
then apply Eqs.~\eqref{eq:39} and \eqref{eq:40} to convert this result
to different densities.  Note also that, because of the translation
and rotation invariance properties, $K$ is expected to be of the form
$K\bigl( \lvert x - \bar x \rvert \bigr)$, i.e.\ it is a function of a
single real argument.  Figure~\ref{fig:3} shows the numerical results
obtained for dimension $n = 2$; again, we can check that all expected
properties for $K$ are indeed satisfied.  For the two-dimensional
Delaunay interpolator we find $\mathcal{N}_\mathrm{eff} \simeq 3.31$.

\subsection{Natural neighbor}
\label{sec:natural-neighbor}

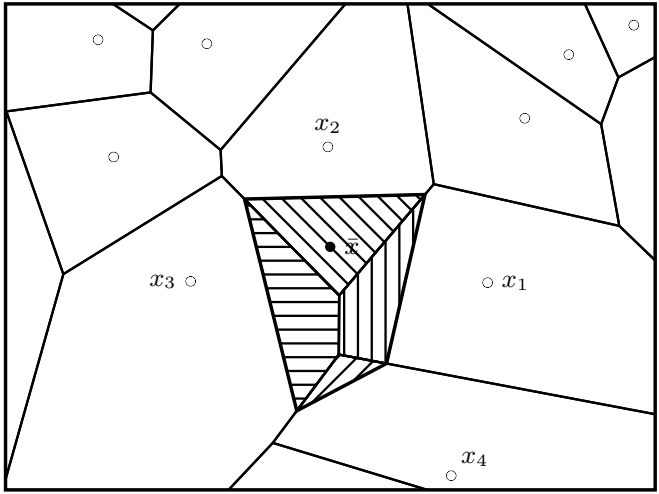
\begin{figure}[!t]
  \begin{center}
    \resizebox{\hsize}{!}{%
      \psset{unit=1cm}
      \begin{pspicture}(-4, -3)(4, 3)
  \psframe[linewidth=1.2pt](-4, -3)(4, 3)
  \begin{psclip}{\psframe[linestyle=none](-4, -3)(4, 3)}
    \psset{dotstyle=o}
    \psdots(-1.51316, 2.49047)
    \psdots(1.47982, -2.80481) \uput[45](1.47982, -2.80481){$x_4$}
    \psdots(3.71583, 2.71800)
    \psdots(-2.32970, 3.31605)
    \psdots(-1.58455, 4.68019)
    \psdots(1.33337, -3.28599)
    \psdots(-2.65509, 1.10293)
    \psdots(4.17089, 1.89607)
    \psdots(2.38058, 1.57684)
    \psdots(2.92183, 2.35638)
    \psdots(-0.0307422, 1.22487) \uput[90](-0.0307422, 1.22487){$x_2$}
    \psdots(1.92673, -0.438251) \uput[0](1.92673, -0.438251){$x_1$}
    \psdots(-2.84451, 2.54006)
    \psdots(-1.70994, -0.421731) \uput[180](-1.70994, -0.421731){$x_3$}
    \psdots(-4.64795, 0.405211)
    \psline(-1.22100, 3.59602)(-2.17368, 2.65378)(-2.20199, 1.89370)(-1.34402, 1.18760)(0.740799, 3.62958)(0.741719, 3.66000)(-1.22100, 3.59602)
    \psline(0.104356, -1.31958)(-0.704878, -2.40275)(14.1005, -6.90894)(6.42024, -2.51230)(0.104356, -1.31958)
    \psline(3.95365, 11.5003)(1.83866, 5.78721)(3.52830, 2.07724)(14.1005, 7.93058)(14.1005, 11.5003)(3.95365, 11.5003)
    \psline(-2.17368, 2.65378)(-1.22100, 3.59602)(-14.6571, 10.9354)(-2.17368, 2.65378)
    \psline(-15.6166, 11.5003)(-15.6166, 11.5003)(-14.6571, 10.9354)(-1.22100, 3.59602)(0.741719, 3.66000)(1.83866, 5.78721)(3.95365, 11.5003)(-15.6166, 11.5003)
    \psline(14.1005, -6.90894)(14.1005, -6.90894)(-0.704878, -2.40275)(-5.19601, -7.17461)(-5.19601, -7.17461)(14.1005, -7.17461)(14.1005, -6.90894)
    \psline(-3.96897, 1.66080)(-3.27061, -0.333922)(-1.32924, 0.869551)(-1.34402, 1.18760)(-2.20199, 1.89370)(-3.96897, 1.66080)
    \psline(14.1005, 7.93058)(3.52830, 2.07724)(3.31714, 1.50424)(3.53951, 0.257164)(6.42024, -2.51230)(14.1005, -6.90894)(14.1005, -6.90894)(14.1005, 7.93058)
    \psline(1.26716, 0.768959)(3.53951, 0.257164)(3.31714, 1.50424)(0.916533, 3.17102)(1.26716, 0.768959)
    \psline(0.741719, 3.66000)(0.740799, 3.62958)(0.916533, 3.17102)(3.31714, 1.50424)(3.52830, 2.07724)(1.83866, 5.78721)(0.741719, 3.66000)
    \psline(-1.34402, 1.18760)(-1.32924, 0.869551)(0.107643, -0.595775)(1.26716, 0.768959)(0.916533, 3.17102)(0.740799, 3.62958)(-1.34402, 1.18760)
    \psline(0.107643, -0.595775)(0.104356, -1.31958)(6.42024, -2.51230)(3.53951, 0.257164)(1.26716, 0.768959)(0.107643, -0.595775)
    \psline(-3.96897, 1.66080)(-2.20199, 1.89370)(-2.17368, 2.65378)(-14.6571, 10.9354)(-15.6166, 11.5003)(-3.96897, 1.66080)
    \psline(-1.32924, 0.869551)(-3.27061, -0.333922)(-5.19601, -7.17461)(-0.704878, -2.40275)(0.104356, -1.31958)(0.107643, -0.595775)(-1.32924, 0.869551)
    \psline(-5.19601, -7.17461)(-5.19601, -7.17461)(-3.27061, -0.333922)(-3.96897, 1.66080)(-15.6166, 11.5003)(-15.6166, 11.5003)(-15.6166, 11.5003)(-15.6166, -7.17461)(-5.19601, -7.17461)
    \psset{dotstyle=*}
    \psdots(0.00000, 0.00000) \uput[0](0,0){$\bar x$}
    \psline[linewidth=1.2pt](-1.05161, 0.586426)(-0.411292, -2.00978)(0.687987, -1.42980)(1.15922, 0.641914)(-1.05161, 0.586426)
    \pspolygon[fillstyle=hlines, hatchangle=0.00000](0.104356, -1.31958)(0.107643, -0.595775)(-1.05161, 0.586426)(-0.411292, -2.00978)
    \pspolygon[fillstyle=hlines, hatchangle=45.0000](0.104356, -1.31958)(-0.411292, -2.00978)(0.687987, -1.42980)
    \pspolygon[fillstyle=hlines, hatchangle=90.0000](0.107643, -0.595775)(0.104356, -1.31958)(0.687987, -1.42980)(1.15922, 0.641914)
    \pspolygon[fillstyle=hlines, hatchangle=135.000](0.107643, -0.595775)(1.15922, 0.641914)(-1.05161, 0.586426)
  \end{psclip}
\end{pspicture}}
    \caption{The evaluation of coefficients for the natural neighbor
      interpolation.  A first Voronoi tessellation for the original
      set of locations (empty circles) is calculated.  Then, the point
      $x$ (filled circle) for which we want to obtain the interpolated
      value is added to the original locations, and a second Voronoi
      tessellation is evaluated.  The new tessellation coincides with
      the old one, with the only difference that the new point $x$ is
      now ``stealing'' the area for its Voronoi cell from neighboring
      cells (the ones corresponding to the points $x_1$, $x_2$, $x_3$,
      and $x_4$ in the case considered in this figure).  The various
      ``stolen areas'' (marked in this figure with parallel lines of
      different orientation) are then used to evaluate the
      coefficients to be inserted in Eq.~\eqref{eq:30}.}
    \label{fig:4}
  \end{center}
\end{figure}

The natural neighbor interpolator \citep{Sibson}, also called area
stealing interpolator, is still related to the Voronoi cells.  Its
construction is carried out in the following way.  First, the Voronoi
cells for the points $\{ x_i \}$ are calculated.  Then, the point $x$
is added to the set, and the new Voronoi cells are calculated.  The
cell corresponding to $x$ in the new configuration overlaps some parts
(stolen areas) of cells originally owned by nearby points: These
points, called \textit{natural neighbors}, will be involved into the
interpolation at $x$.  Specifically, the value assigned to $x$ will be
the weighted average [see Eq.~\eqref{eq:30}] of values at the natural
neighbors, with weights equal to the overlap areas (see
Fig.~\ref{fig:4}).

The natural neighbor interpolator is linear, normalized, invariant
upon translation, rotation, and scaling, exact, bounded, and weakly
local.  All these properties are easily verified except, perhaps, the
exact property.  Suppose that the point $x$ moves toward one of the
locations, say $x_3$.  As $x$ get closer and closer to $x_3$, the
stolen area will almost be entirely inside the (old) cell
corresponding to $x_3$.  In the limit where $x$ coincides with $x_3$,
the stolen area will be the intersection of the cell of $x_3$ with the
half-plane $\bigl\{ \bar x \bigm| \bigl\langle x - x_3, \bar x
\bigr\rangle > 0 \bigr\}$, where $\langle \cdot, \cdot \rangle$
denotes the scalar product.

Again, analytical calculations to obtain $K$ for this smoothing
technique are feasible only in dimension $n = 1$.  In this case,
surprisingly, the natural neighbor smoothing is totally equivalent to
the Delaunay triangulation, and thus Eq.~\eqref{eq:63} holds.  

For higher dimensions, we can only obtain numerical estimates for
$K\bigl( \lvert x - \bar x \rvert \bigr)$.  Figure~\ref{fig:3} shows
these estimates in dimension $n = 2$; in this case we obtain
$\mathcal{N}_\mathrm{eff} \simeq 3.41$.

\subsection{Moving weights}
\label{sec:moving-weights}

One of the most common interpolating techniques used in Astronomy is
based on a simplified form of Eq.~\eqref{eq:30} in which the $i$-th
weight depends only on $x$ and on the $i$-th location:
\begin{equation}
  \label{eq:65}
  S\bigl( x; \bigl\{ (x_i, y_i) \bigr\}\bigr) = \biggl[ \sum_{i=1}^N
  w(x - x_i) y_i \biggr] \biggm/ \biggl[ \sum_{i=1}^N w(x - x_i)
  \biggr] \; .
\end{equation}
The function $w(x - x_i)$, taken here to be non-negative, is usually
chosen to have a peak at $x = x_i$ and to decrease to zero as $\lvert
x - x_i \rvert$ increases.  This way, the estimated value at $x$ will
be basically determined by the neighboring points.  Often the weight
function is isotropic, i.e.\ $w(x) = w\bigl( \lvert x \rvert \bigr)$.
Commonly used weight functions include Gaussians, top-hats, and
inverse distances of the form $w(x) = 1/\lvert x \rvert^\alpha$ with $\alpha$
fixed [this is precisely our toy-interpolator defined in
Eq.~\eqref{eq:2}].

Some of the properties of the moving weight interpolator strongly
depend on the weight function used.  In general, this interpolator is
linear, normalized, invariant upon translations, and bounded.  Hence,
the kernel associated to the smoothing satisfies $K(x - \bar x) \le 1$
and is normalized to unit.  The interpolator is invariant upon
rotation if and only if $w$ is isotropic.  If $w$ is homogeneous of
degree $g$ (as for example for inverse distances weights), i.e.\ if
$w(k x) = k^g w(x)$, then the interpolator is invariant upon scaling .
The interpolator is generally not exact; however, if $w(x)$ is bounded
for $x \ne 0$ and goes to infinity for $x = 0$ (e.g., in case of
inverse distance weights), then the interpolator is exact and thus we
expect $K(0) = 1$.  The interpolator is strongly local if and only
if $w(x)$ has compact support, in which case $K$ has also compact
support; it is otherwise global.  Finally, we observe that this
interpolator is defined if just one point is inside the support of
$w$.  Hence, if $w$ has compact support of area $a$, we have $P_0(x) =
\exp [ - \rho a]$; in other cases $P_0(x)$ vanishes.

The statistical properties of this smoothing have been studied in
detain in a separate paper \citep{2001A&A...373..359L}.  There we have
shown that the kernel can be evaluated using the set of equations
\begin{align}
  \label{eq:66}
  Q(s) & {} = \int_X \left[ \e^{-s w(x)} - 1\right] \, \diff x \; ,
  \\
  \label{eq:67}
  K(x) & {} = \frac{w(x)}{1 - P_0} \int_0^\infty \e^{-w(x) s +
  \rho Q(s)} \, \diff s \; .
\end{align}
In the same paper we have also explicitly shown several of properties
stated above for this interpolator.  In addition, we have shown that
the kernel $K$ associated with this smoothing closely resemble the
weight function $w$ used, but is generally larger.  For example, if we
define the area $\mathcal{N}$ of $w$ analogously to Eq.~\eqref{eq:48},
we find $\mathcal{N}_\mathrm{eff} \ge \mathcal{N}$.  The kernel $K$
however converges to $w$ as $\rho \rightarrow \infty$.

\subsection{Fixed weights}
\label{sec:fixed-weights}

For some studies a non-normalized form of the smoothing \eqref{eq:65}
is still used:
\begin{equation}
  \label{eq:68}
  S\bigl( x; \bigl\{ (x_i, y_i) \bigr\} \bigr) = \sum_{i=1}^N w(x -
  x_i) y_i \; .
\end{equation}
This smoothing has been employed, e.g., in some weak lensing studies
\citep{KS, 1995ApJ...446L..55T, 1997ApJ...475...20L,
  1999AJ....117.2024F}.

The interpolation procedure \eqref{eq:68} lacks several of the
properties described above.  In particular, it is only linear and
invariant upon translation.  If $w(x) = w \bigl( \lvert x \rvert
\bigr)$ is isotropic, then the rotation invariance holds.  The
interpolator is \textit{not\/} normalized (which is quite unusual for
a smoothing technique); however, if the weight function $w$ is
normalized to $1$, i.e.
\begin{equation}
  \label{eq:69}
  \rho \int_X w(x) \, \diff x = 1 \; ,
\end{equation}
then Eq.~\eqref{eq:28} still holds.  The interpolator is not exact and
not bounded; it is strongly local if $w$ has compact support,
otherwise it is global; it is always defined, so that $P_0(x) =
P_0(x,x') = 0$.

A statistical analysis of this interpolator is straightforward.  The
result obtained is that the kernel $K$ coincides with the weight
function $w$.  As expected, the kernel is not necessary bounded
superiorly and is not necessary equal to unit at $x = 0$ (see
\citealp{LPM} for a detailed discussion of this smoothing technique
applied to weak lensing data).

\subsection{Other interpolation techniques}
\label{sec:other-interp-techn}

Clearly, a complete analysis of all smoothing methods currently used
would be impossible here; on the other hand, the general
classification technique described in Sect.~\ref{sec:classifications}
should allow one to characterize other interpolators.  We want to
mention however a couple of interesting smoothing techniques, namely
the splines and the kriging.

Splines are an interesting and complex topic (see \citealp{NR} for a
short introduction; a complete discussion can be found in
\citealp{Boor}).  They find several applications in many different
fields in Astronomy, and can also be used for ``cosmetic
treatments.''\@ Splines come in several variants with slightly
different properties.  For example, some variants of splines are exact
interpolators, while others are not.  Another peculiarity is that
often splines are \textit{not\/} bounded interpolators.  In order to
avoid any possible confusion among the different variants of splines,
we prefer here not to consider this smoothing technique in detail.  We
stress again, however, that a statistical analysis of any particular
variant can be carried out without difficulties using the framework
described in this paper.

Kriging is another interesting interpolation method.  Although it is
seldom used in Astronomy, it finds several applications in geophysics
(see, however, \citealp{1991ApJ...378..106A} for an example of
astronomical application of kriging).  A complete discussion of this
interpolation method, which also has some variants, is beyond the
scope of this paper.  Here, we just observe that kriging is a
non-linear method, and thus cannot profitably be analyzed with the
techniques discussed in this article.

\section{Conclusions}
\label{sec:conclusions}

In this paper we have considered the statistical properties of
interpolation techniques.  The discussion has originally been kept at
a high level of abstraction, and this has given us the ability to
characterize smoothing methods in terms of simple properties.  In
particular, we have made a classification of interpolation techniques
and we have derived several statistical results associated with each
class of interpolators.  A comparison of this analysis with a more
technical one carried out in a separate paper for a specific smoothing
method \citep{2001A&A...373..359L} clearly shows the advantages of
using an abstract approach to the problem.  In the second part of this
paper we have applied the results obtained to several commonly used
smoothing methods.

\acknowledgements{I would like to thank Peter Schneider and Arlie
  Petters for many helpful discussions and the Referee, Stephane
  Colombi, for several interesting suggestions.  This work has been
  partially supported by the Deutsche Forschungsgemeinschaft and by
  the TMR Network ``Gravitational Lensing: New Constraints on
  Cosmology and the Distribution of Dark Matter'' of the EC under
  contract No.~ERBFMRX-CT97-0172.}

\bibliographystyle{aa}
\bibliography{../lens-refs}

\end{document}